\documentclass{article}

\usepackage{amssymb,amsmath,mathtools,xcolor,graphicx,xspace,colortbl,ragged2e,rotating} %
\usepackage{amssymb,amsmath,mathtools,xcolor,graphicx,xspace,colortbl,ragged2e,rotating}  
\usepackage{booktabs}
\usepackage{jheppub}
\graphicspath{{Conformal_renormalization of Einstein-AdS gravity_graphics/}{Conformal_renormalization of Einstein-AdS gravity_tcache/}{Conformal_renormalization of Einstein-AdS gravity_gcache/}}
\DeclareGraphicsExtensions{.pdf,.eps,.ps,.png,.jpg,.jpeg}
\graphicspath{{Conformal_renormalization of Einstein-AdS gravity_graphics/}{Conformal_renormalization of Einstein-AdS gravity_tcache/}{Conformal_renormalization of Einstein-AdS gravity_gcache/}}
\DeclareGraphicsExtensions{.pdf,.eps,.ps,.png,.jpg,.jpeg}

\allowdisplaybreaks

\begin{document}

\title{Holographic Counterterms from Restricted Weyl Symmetry in Even-Dimensional Einstein–AdS Gravity}

\author{Giorgos Anastasiou}

\emailAdd{georgios.anastasiou@uai.cl}

\affiliation{Universidad Adolfo Iba\~nez, Facultad de Artes Liberales, Departamento de Ciencias,\\ Av. Diagonal Las Torres, 2640 Pe\~nalolen, Chile }

\abstract{We develop a symmetry-based approach to renormalization for even-dimensional Einstein–AdS gravity based on infinitesimal bulk Weyl transformations that preserve its field equations asymptotically. We determine a bulk-covariant completion of the Einstein–Hilbert density by requiring its variation under restricted Weyl transformations with parameter $\Sigma(X)$ to vanish modulo terms whose integrals remain finite as the radial cutoff is removed. This requirement induces two types of compensating terms with fixed couplings: the Euler density and a hierarchy of covariant total derivatives. The resulting recursive relations admit a factorized solution in terms of shifted Laplacians, yielding a closed-form completion in arbitrary even bulk dimensions. On Einstein backgrounds, these expressions agree, after matching normalizations, with those obtained through the ambient construction of conformal invariants. The associated renormalized-volume identity establishes on-shell finiteness for Euclidean AlAdS Einstein geometries in arbitrary even dimensions. The prescription reproduces the known conformally renormalized actions in four, six and eight dimensions and provides an explicit ten-dimensional completion involving powers of the Laplacian up to $\Box^3$. Applications to eight- and ten-dimensional AdS black holes with non-conformally-flat Einstein horizons yield finite on-shell actions in agreement with Holographic Renormalization.}

\maketitle

\section{Introduction}
\label{Intro}

The anti-de Sitter/Conformal Field Theory (AdS/CFT) correspondence provides a concrete realization of holographic duality by relating a gravitational theory in an AdS$_{d+1}$ spacetime to a CFT$_d$ defined on its boundary~\cite{Maldacena:1997re,Gubser:1998bc,Witten:1998zw}. At the level of generating functionals, the correspondence is expressed schematically as the equality of the partition functions on the two sides of the duality
\begin{equation}Z_{CFT}\left (\phi _{\left (0\right )}\right ) =Z_{AdS}\left (\phi  \rightarrow \phi _{\left (0\right )}\right ) \,,
\end{equation}
where $\phi _{\left (0\right )}$ denotes the fixed value of the bulk field $\phi$ at the boundary. In a regime admitting a semiclassical bulk description, the leading contribution to the gravitational partition function is obtained from the saddle-point approximation, 
\begin{equation}
Z_{CFT}\left (\phi _{\left (0\right )}\right ) \approx \exp \left (-I_{\rm{E}}\right )_{\phi  \rightarrow \phi _{\left (0\right )}} \,,
\end{equation}
where $I_{\rm{E}}$ is the Euclidean gravitational action evaluated on a classical solution whose value at the conformal boundary is $\phi _{\left (0\right )}$.

A fundamental difficulty is that the on-shell action  is infrared (IR) divergent due to the infinite volume of asymptotically locally AdS (AlAdS) spacetimes. In $D=d+1$ dimensions, metrics in this class admit a Fefferman--Graham (FG) expansion~\cite{AST_1985__S131__95_0,fefferman2012ambient} as
\begin{align}
ds^{2} &= g_{\mu \nu} dX^{\mu} dX^{\nu}=\frac{\ell^2}{z^2} \left(dz^2 + \gamma_{ij} \left(z,x\right) dx^{i} dx^{j}\right) \,, \label{Gaussnormal}\\
\gamma_{ij} \left(z,x\right) &= g_{\left(0\right) ij} \left(x\right)+ \frac{z}{\ell} g_{\left(1\right) ij} \left(x\right) + \frac{z^2}{\ell^2} g_{\left(2\right) ij} + \ldots +\frac{z^{d}}{\ell^d}\left(\,g_{\left(d\right)ij}+h_{\left(d\right)ij} \log \frac{z^2}{\ell^2} \right)+\ldots \,,
\label{FGexpansion}
\end{align}
where $z$ is the radial coordinate, $\ell$ is the AdS radius and the Latin indices label coordinates on the constant-$z$ hypersurfaces. The logarithmic term appears when the bulk dimension is odd and, in the holographic setting, is associated with the Weyl anomaly~\cite{Duff:1993wm,Henningson:1998gx,deHaro:2000vlm}. 

For Einstein-AdS (EAdS) gravity, the odd-power coefficients preceding the normalizable order vanish, while the  coefficients $g_{\left(k\right)ij}$ with $k<d$ are determined locally by the equations of motion (EOM) as functions of the boundary metric $g_{\left(0\right)ij}$. In particular, the first two subleading terms of the FG expansion are
\begin{align}
g_{\left(2\right)ij} &= - \ell^2 \mathcal{S}_{\left(0\right)ij} \,, \label{g2}\\
g_{\left(4\right)ij} &= \frac{\ell^4}{4(d-4)}\,\left[(d-4)\mathcal{S}_{\left(0\right)ik}\mathcal{S}^k_{\left(0\right)j}-\mathcal{B}_{\left(0\right)ij}\right] \,.
\label{g4}
\end{align}
Here
\begin{align}
\mathcal{S}_{\left(0\right)ij}&=\frac{1}{d-2}\left( \mathcal{R}_{\left(0\right)ij}-\frac{\mathcal{R}_{\left(0\right)}}{2(d-1)}\,g_{\left(0\right)ij}\right) \,,  \label{Schoutenboundary} \\
\mathcal{B}_{\left(0\right)ij}&=\mathcal{D}^{k}_{\left(0\right)}\mathcal{C}_{\left(0\right)ijk} + \mathcal{S}_{\left(0\right)}^{kl}\mathcal{W}_{\left(0\right)ikjl} \,, \label{Bach} \\
\mathcal{C}_{\left(0\right)ijk} &=\mathcal{D}_{\left(0\right)k}\mathcal{S}_{\left(0\right)ij}-\mathcal{D}_{\left(0\right)j}\mathcal{S}_{\left(0\right)ik} \,,\label{cotton}
\end{align}
denote the boundary Schouten, Bach and Cotton tensors, respectively. The conformal boundary lies at $z=0$, where the second-order pole of the bulk metric induces the asymptotic conformal structure and, at the same time, gives rise to the IR divergences of the gravitational action. These are the holographic counterpart of ultraviolet (UV) divergences in the dual field theory and must be removed before any holographic observables, conserved charges or thermodynamic potentials can be defined.

The standard framework for rendering the gravitational action finite is Holographic Renormalization (HR)~\cite{Henningson:1998gx,deHaro:2000vlm,Skenderis:2002wp}. In this approach, the bulk fields are expanded asymptotically in FG form and a radial cutoff $\epsilon$ is introduced to regulate the action. Its divergent contributions are then cancelled by local covariant counterterms constructed from the induced fields at the cutoff hypersurface. Due to this prescription, the renormalized EAdS action is given by
\begin{equation}
I_{\text{ren}} =
\frac{1}{16 \pi G_N}
\int\limits_{\mathcal{M}_{d+1}}
d^{d+1}x\sqrt{\vert g \vert} \left(R-2\Lambda\right) 
+\frac{1}{8 \pi G_N}\int\limits_{\partial \mathcal{M}_{d+1}}
d^{d}x\,\sqrt{\vert h \vert}\,K 
+\int\limits_{\partial \mathcal{M}_{d+1}}d^{d}x\,\mathcal{L}_{ct}
\,,
\label{Sct}
\end{equation}
where $\Lambda=-\frac{\left(D-1\right) \left(D-2\right)}{2 \ell^2}$ is the cosmological constant, $h_{ij}=\frac{\ell^2}{z^2}\,\gamma_{ij}$ is the induced metric at constant-$z$, $K$ is the trace of the extrinsic curvature
\begin{equation}
    K_{ij}=-\dfrac{1}{2N}\,\partial_{z}h_{ij}\,, \label{K}
\end{equation}
with $N=\frac{\ell}{z}$ denoting the lapse function. The counterterms can be written as
\begin{align}\label{L-HRCounter}
8\pi G_N  \mathcal{L}_{ct} &= -\frac{d-1}{\ell }\sqrt{\vert h \vert}-\frac{\ell \left(d-1\right)\sqrt{\vert h \vert}}{%
\left(d-2\right)}\mathcal{S} + \frac{\ell^3 \sqrt{\vert h \vert}}{2\left(d-4\right)} \delta^{ik}_{jl}\mathcal{S}^{j}_{i}\mathcal{S}^{l}_{k} \notag \\
&- \frac{\ell^{5}\sqrt{\vert h \vert}}{\left(d-2\right)\left(d-4\right)\left(d-6\right)} \left(\mathcal{S}^i_j \mathcal{B}^j_i +\frac{d-4}{2} \delta^{ikm}_{jln}\mathcal{S}^{j}_{i}\mathcal{S}^{l}_{k} \mathcal{S}^{n}_{m} \right) +\mathcal{O}\left(\mathcal{R}^4\right)\,,
\end{align}
up to cubic terms in the curvature, where $\mathcal{S}=h^{ij}\mathcal{S}_{ij}$. Here, $\mathcal{S}_{ij}$ and $\mathcal{B}_{ij}$ are the Schouten and Bach tensors of the induced metric $h_{ij}$, respectively. The counterterms were originally derived in Refs.~\cite{Henningson:1998gx,deHaro:2000vlm,Emparan:1999pm,Balasubramanian:1999re} but here we present them in a basis spanned by fundamental tensors of conformal geometry~\cite{Boulanger:2004eh}. One may derive the same counterterm series by imposing Dirichlet boundary conditions for the conformal boundary metric $g_{\left(0\right) ij}$ ~\cite{Papadimitriou:2005ii}. The problem then reduces to the solution of a radial Hamilton-Jacobi equation and leads to a recursive relation that involves the expansion of the canonical momentum. The resulting renormalized action~\eqref{Sct} determines the boundary stress tensor and, for even boundary dimension, the conformal anomaly~\cite{Deser:1993yx,Henningson:1998gx}.

Although HR is systematic, the explicit form of the counterterm series becomes increasingly involved as the bulk dimension grows. This has motivated alternative prescriptions in which the same asymptotic information is reorganized in a manifestly bulk-covariant form. Of particular relevance to the present work is the observation that Weyl invariant gravitational functionals exhibit well-behaved asymptotics in the presence of conformally compact boundaries. Concrete examples occur in four bulk dimensions, both for gravitational actions \cite{Grumiller:2013mxa} and for codimension-2 geometric functionals \cite{Anastasiou:2025dex}. This suggests that bulk Weyl symmetry may provide a fundamental principle for the cancellation of the IR divergences.

The connection between bulk Weyl invariance and infrared finiteness can be understood from the asymptotic behavior of the Weyl tensor. For generic AlAdS geometries, its independent components $W^{ij}_{ms}$, $W^{iz}_{ms}$ and $W^{iz}_{jz}$ fall off at least quadratically in the radial coordinate~$\mathcal{O} \left(z^2\right)$, as detailed in Appendix~\ref{GC}. Consequently, in even bulk dimension $D=2n$, a scalar polynomial containing $n$ Weyl tensors behaves as $O(z^{2n})$, compensating the leading $z^{-2n}$ divergence of the bulk volume density. Denoting schematically by $f\left(W^n\right)$ an arbitrary degree-$n$ contraction of Weyl tensors, the corresponding bulk integral scales as
\begin{equation}
\int \limits_{\mathcal{M}_{2n}} d{^{2n}X}\sqrt{\vert g \vert} \, f\left(W^n\right) \sim \int_{z \ge \epsilon} d{z} \int \limits_{\partial\mathcal{M}_{2n}} d{^{2n-1}x} \frac{\sqrt{\vert g_{\left(0\right)} \vert}\ell^{2n}}{z^{2n}} z^{2n} = I_\text{fin}+ \mathcal{O} \left(\epsilon\right) \,.
\label{pfaffian}
\end{equation}
Hence, this action is free of IR divergences. This simple power-counting argument makes manifest why polynomial Weyl invariants have sufficiently soft asymptotic behavior to define finite bulk functionals on generic AlAdS geometries. Conformal invariants involving covariant derivatives require a separate asymptotic analysis, as illustrated by the six- and eight-dimensional constructions discussed below~\cite{Anastasiou:2020mik,Anastasiou:2026jrt}. This observation motivates seeking a bulk-covariant renormalization prescription organized by Weyl transformation properties.

Conformal Renormalization (CR) is a concrete realization of this connection that relies on embedding Einstein--AdS gravity into Conformal Gravity (CG). In four dimensions, the CG action is IR finite, as suggested by the power-counting argument above and demonstrated explicitly in Ref.~\cite{Grumiller:2013mxa}. Its solution space contains Einstein metrics, which can be selected by imposing Neumann boundary conditions~\cite{Maldacena:2011mk}. On this sector, the CG action reduces to the Einstein--Hilbert action supplemented by the Gauss--Bonnet invariant with a fixed coupling. The resulting action inherits the finiteness of the parent action and admits a Dirichlet variational principle for the boundary metric $g_{\left(0\right)ij}$~\cite{Anastasiou:2016jix}, which is the source of the stress tensor of the dual CFT.

Extending CR to higher dimensions requires identifying combinations of local conformal invariants whose equations of motion admit the Einstein sector. This is a nontrivial task, particularly because of the increasing number and complexity of conformal invariants involving covariant derivatives. The local conformal invariants have been classified in six~\cite{parker1987invariants,Erdmenger:1997gy} and eight dimensions~\cite{Boulanger:2004zf,Boulanger_2022}, while the corresponding combinations admitting an Einstein sector were identified in Refs.~\cite{Lu:2013hx} and~\cite{Boulanger:2025oli}, respectively. In both cases, CR selects the Einstein sector through generalized Neumann boundary conditions, yielding bulk-covariant counterterms that render the in Einstein--AdS action finite and ensure agreement with holographic renormalization~\cite{Anastasiou:2020mik,Anastasiou:2026jrt}. The resulting bulk densities can be written as linear combinations of scalar invariants formed from fully antisymmetrized products of the Riemannian part of the AdS curvature tensor,
\begin{equation}
F_{\mu\nu}^{\alpha\beta}
=
R_{\mu\nu}^{\alpha\beta}
+
\frac{1}{\ell^2}\delta_{\mu\nu}^{\alpha\beta}\,,
\label{Ftensor}
\end{equation}
and Laplacian descendants of these invariants. The connection between IR finiteness and Weyl symmetry thus identifies the AdS curvature tensor $F_{\mu\nu}^{\alpha\beta}$ as a natural building block of the renormalized EAdS action. This has a direct geometric basis: on EAdS spacetimes, $F_{\mu\nu}^{\alpha\beta}$ coincides with the Weyl tensor ($F^{\alpha \beta}_{\mu \nu}\vert_E= W^{\alpha \beta}_{\mu \nu} $).

The asymptotic structure of Einstein--AdS gravity suggests that full off-shell Weyl invariance may be stronger than necessary to determine the gravitational counterterms. In particular, the divergent contributions to the regulated action are determined by the non-normalizable FG coefficients, which are locally constrained by the Einstein--AdS equations, whereas the independent normalizable data encode state-dependent information~\cite{deHaro:2000vlm,Skenderis:2002wp}.
We therefore introduce a renormalization scheme based on infinitesimal bulk Weyl transformations that preserve the EAdS equations asymptotically. The boundary metric and the individual FG coefficients may vary, provided their relations imposed by the EOM remain satisfied. We require the Einstein–Hilbert density, supplemented by the Euler invariant and covariant total derivatives, to be Weyl invariant up to terms whose integrals remain finite as the cutoff is removed.

Under this class of restricted Weyl transformations, $F_{\mu\nu}^{\alpha\beta}$ transforms homogeneously through the non-normalizable orders. Applying the completion criterion to the Einstein--Hilbert density supplemented by the Euler invariant fixes its coupling, recovering the topologically renormalized Einstein--AdS action~\cite{Olea:2005gb,Miskovic:2009bm,Miskovic:2014zja}. While this completion is sufficient in four dimensions~\cite{Miskovic:2009bm,Miskovic:2014zja} and for asymptotically conformally flat geometries (ACF) in higher even dimensions~\cite{Anastasiou:2020zwc}, nonvanishing boundary Weyl curvature generates additional obstructions. We show that these obstructions can be compensated by covariant total derivatives constructed from powers of the Laplacian acted on scalar polynomials in the $F_{\mu\nu}^{\alpha\beta}$. For each scalar $\Phi_{\left(i\right)}$ constructed from the fully antisymmetrized product of $i$ AdS curvature tensors
\begin{equation}
\Phi _{\left (i\right )} =\delta _{\mu _{1}\ldots \mu _{2i}}^{\nu _{1}\ldots \nu _{2i}}F_{\nu _{1}\nu _{2}}^{\mu _{1}\mu _{2}} \cdots F_{\nu _{2i -1}\nu _{2i}}^{\mu _{2i -1}\mu _{2i}} \,,
\label{phiF}
\end{equation}
the Laplacian descendants can be combined into expressions with definite restricted Weyl weights. These combinations factorize into products of shifted Laplacians, and their associated densities have vanishing restricted Weyl weight at the critical Laplacian order $q_{\mathrm{crit}}=n-i$ in $D=2n$. This yields a closed-form expression for the Weyl-completed density in arbitrary even dimensions, without requiring the classification of higher-dimensional off-shell conformal invariants.

On Einstein backgrounds, this structure agrees with the ambient construction of Case et al.~\cite{Case:2024renormalized}, which relates convergent conformal curvature integrals to the renormalized volume and Euler characteristic. Here, we recover the corresponding density from a restricted bulk Weyl criterion that fixes the Euler coupling and derivative corrections without imposing infrared finiteness as an independent condition.

The resulting closed-form expression reproduces the known conformally renormalized Einstein--AdS actions in four, six and eight dimensions~\cite{Anastasiou:2016jix,Anastasiou:2020mik,Anastasiou:2026jrt} and yields an explicit ten-dimensional completion involving derivatives up to $\Box^3\Phi_{\left(2\right)}$. As an independent consistency check, we evaluate the eight- and ten-dimensional actions on Schwarzschild--AdS black holes with non-conformally-flat boundary geometries. Despite the nontrivial dependence of the individual curvature invariants on the horizon geometry, the resulting actions are finite and agree with the standard holographic counterterm prescription. These results demonstrate that restricted Weyl completion provides an alternative symmetry-based construction of gravitational counterterms, with infrared finiteness verified independently of the criterion used to determine their coefficients.

The paper is organized as follows. Section~\ref{weyleinsteinorbit} introduces the restricted Weyl transformations and derives their geometric conditions and FG realization. Section~\ref{TopologicalRenormalization} determines the Euler-density coupling and identifies the residual curvature-dependent obstruction. Section~\ref{6D} develops the six-dimensional completion. Section~\ref{evenD} derives the recursive identities and their factorized solution, obtains the completed density in arbitrary even dimensions, and specializes the result to eight and ten dimensions. Section~\ref{topologicalBH} examines the resulting actions on AdS black holes with non-conformally-flat Einstein horizons and compares their finite values with standard holographic renormalization. Finally, Sec.~\ref{discussion} discusses the implications and possible extensions of the framework.

\section{Restricted Weyl transformations}
\label{weyleinsteinorbit}

As discussed in Sec.~\ref{Intro}, bulk Weyl symmetry can organize gravitational densities with sufficiently soft asymptotic behavior to yield functionals free of IR divergences. This suggests that Weyl transformation properties can be used to determine the local counterterms required to cancel the divergences of the EAdS action. These divergences are governed by the non-normalizable FG data, whose subleading coefficients are locally constrained by the EOM. A generic infinitesimal bulk Weyl rescaling,
\begin{equation}
\delta_\sigma g_{\mu\nu}=2\sigma(X)g_{\mu\nu}\,,
\end{equation}
however, does not preserve the EAdS equations. In CG, for example, Weyl-related solutions can include both Einstein and non-Einstein geometries~\cite{Kozameh:1985jv}. Full off-shell Weyl invariance may therefore impose a stronger requirement than is necessary for determining the gravitational counterterms.

We instead seek infinitesimal Weyl transformations that preserve the EAdS equations. We therefore restrict the scaling parameter $\sigma(X)$ to admissible functions $\Sigma(X)$ satisfying
\begin{equation}
\delta_\Sigma g_{\mu\nu}^{(E)}=2\Sigma g_{\mu\nu}^{(E)} \,,
\qquad
\delta_\Sigma\mathcal E_{\mu\nu}[g^{(E)}]\simeq0 \,,
\label{asymptEAdS}
\end{equation}
where $\mathcal E_{\mu\nu}=R_{\mu\nu}+(D-1)\ell^{-2}g_{\mu\nu}$. Here, $\simeq$ denotes equality through the non-normalizable FG orders, leaving contributions at normalizable order unconstrained.

The central claim of our prescription is that rendering the Einstein-Hilbert action invariant under infinitesimal local Weyl rescaling with scaling parameter $\Sigma(X)$ is a sufficient criterion for determining bulk-covariant counterterms in $D=2n$ dimensions. Specifically, we seek the bulk-covariant completion of the Einstein--Hilbert density requiring that
\begin{equation}
\left.\delta_\Sigma\mathcal L^{(2n)}\right|_E\doteq0 \,,
\label{variationdiv}
\end{equation}
where $\doteq$ denotes equality modulo asymptotically integrable contributions. To preserve the bulk Einstein equations, we take the compensating terms $\mathcal L_{\mathrm{weyl}}$ to consist of the Euler invariant and covariant total derivatives such that
\begin{equation}
\mathcal L^{(2n)}
=\sqrt{|g|}\left(R-2\Lambda+\mathcal L_{\mathrm{weyl}}\right) \,.
\end{equation}
The corresponding gravitational action obtains the form
\begin{equation}
I^{(2n)}=\frac{1}{16\pi G_N}
\int_{\mathcal M_{2n}}d^{2n}X\,\mathcal L^{(2n)}
+I_0^{(2n)} \,,
\label{I2naction}
\end{equation}
with the constant $I_0^{(2n)}$ specified in Sec.~\ref{TopologicalRenormalization}. The asymptotic integrability criterion ensures that the restricted Weyl variation of the regulated action has no divergent contribution when the cutoff is removed,
\begin{equation}
\left[\delta_\Sigma I^{(2n)}(\epsilon)\right]_{\mathrm{div}}=0 \,,
\end{equation}
where $\epsilon$ is the IR cutoff in FG coordinates.
This does not by itself establish infrared finiteness of the action, which is examined independently in the following sections.

\subsection{Restricted Weyl condition and geometric identities}
\label{restrictedweyl}

We now derive the differential equation governing the restricted Weyl parameter $\Sigma(X)$. To do so, we introduce the infinitesimal Weyl variation of the Ricci tensor that reads
\begin{equation}
\delta_\sigma R_{\mu\nu}
=-(D-2)\nabla_\mu\nabla_\nu\sigma
-g_{\mu\nu}\Box\sigma \,.
\end{equation}
Substituting this relation into Eq.~\eqref{asymptEAdS}, we determine the variation of the Einstein--AdS equations along $\Sigma$, that gives
\begin{equation}
\delta_\Sigma \mathcal{E}_{\mu \nu}=-\left(D-2\right) \nabla_{\mu} \nabla_{\nu} \Sigma -g_{\mu \nu}^{\left(E\right)} \Box \Sigma + \frac{2 \left(D-1\right)}{\ell^2} \Sigma g_{\mu \nu}^{\left(E\right)}\simeq 0 \,.
\end{equation}
Taking the trace of this equation implies that
\begin{equation}
\Box \Sigma -\frac{D}{\ell^2} \Sigma \simeq 0 \,.
\end{equation}
Substituting this relation back into the variation of the EOM yields the restricted Weyl condition,
\begin{equation}
\nabla _{\mu } \nabla_{\nu }\Sigma-\frac{\Sigma }{\ell ^{2}} g_{\mu \nu}^{\left(E\right)}\simeq 0 \,.
\label{weyleinsteincondition}
\end{equation}
The remainder of this equation is $\mathcal{O}(z^{d-2})$ with both indices lowered, or $\mathcal{O}(z^d)$ with one index raised, i.e, the corresponding normalizable orders. These estimates specify the asymptotic accuracy of the restricted identities used below. The significance of this condition is that it provides the geometric identities underlying the counterterm construction. 

In particular, the AdS curvature tensor $F_{\mu \nu }^{\alpha \beta }$ admits non-homogeneous contributions under arbitrary Weyl rescalings, as it can be deduced due to the presence of the Riemann tensor. However, it transforms homogeneously under the restricted Weyl variations~\eqref{weyleinsteincondition}
\begin{equation}
\delta _{\Sigma }F_{\mu \nu }^{\alpha \beta }  
\simeq -2\Sigma F_{\mu \nu }^{\alpha \beta }\,,
\label{weylvarF2}
\end{equation}
as shown explicitly in Appendix~\ref{Weyleinsteincondition2}. Namely, $F_{\mu \nu }^{\alpha \beta }$ transforms along $\Sigma (X)$ with the same scaling weight as the Weyl tensor. An additional identity follows from the integrability condition of Eq.~\eqref{weyleinsteincondition}. Commuting two covariant derivatives acting on $\nabla^{\alpha} \Sigma$ gives
\begin{equation}
R^{\alpha \nu}_{ \lambda \mu} \nabla_{\nu} \Sigma  =\left[\nabla_\lambda,\nabla_\mu \right]\nabla^{\alpha}\Sigma\,.
\end{equation}
Using Eq.~\eqref{weyleinsteincondition}, this relation becomes
\begin{equation}
\left[R^{\alpha \nu}_{ \lambda \mu} -\frac{1}{\ell^2}\left(\delta^{\alpha}_{\mu} \delta^{\nu}_{\lambda} -\delta^{\alpha}_{\lambda}  \delta^{\nu}_{\mu}\right)\right]\nabla_{\nu} \Sigma \simeq 0\,,
\end{equation}
or in terms of the AdS curvature tensor
\begin{equation}
F^{\alpha \nu}_{ \lambda \mu} \nabla_{\nu} \Sigma \simeq 0 \,.
\label{FcottonEinstein}
\end{equation}
Both identities are fundamental to determine the scaling weight of $\Phi_{\left(i\right)}$ and its Laplacian descendants.

\subsection{Fefferman–Graham realization of the restricted Weyl condition}
\label{FGweyl}

We now examine the conformal structure that the scaling parameter $\Sigma$ induces near the conformal boundary. To do so, we need to write Eq.~\eqref{weyleinsteincondition} in FG coordinates. We restrict the analysis to even bulk dimensions, leaving the logarithmic sector of odd-dimensional gravity for future investigation.

Using the Christoffel symbols collected in Appendix~\ref{GC}, the $zz$, $zi$, and $ij$ components of Eq.~\eqref{weyleinsteincondition} can be analyzed independently.
Starting from the $zz$ component, we get
\begin{equation}
z^2 \partial_z^2 \Sigma + z \partial_z \Sigma- \Sigma \simeq 0\,.
\label{Hessianzz}
\end{equation}
Substituting the Frobenius ansatz
\begin{equation}
\Sigma \left(z,x\right)= \sum_{p} z^p a_{\left(p\right)} \left(x\right) \,,
\label{Frobenius}
\end{equation}
into Eq.~\eqref{Hessianzz} yields
\begin{equation}
\left(z^2 \partial_z^2 + z \partial_z - 1\right) z^p a_{\left(p\right)} = \left(p^2-1\right) z^p a_{\left(p\right)} \,,
\end{equation}
which determines the roots $p=\pm1$. Identifying $a_{-1}=\sigma_{\left(0\right)}$ and $a_1=\sigma_{\left(2\right)}$, the radial dependence of $\Sigma$ asymptotically reads
\begin{equation}
\Sigma \left(z,x\right) = \frac{1}{z} \sigma_{\left(0\right)} + z \sigma_{\left(2\right)} + z^{d}\Sigma_N (z,x) \,,
\label{Sigmaexp}
\end{equation}
where $\Sigma_N$ remains undetermined from the $zz$ component by Eq.~\eqref{weyleinsteincondition}. This term is associated with the normalizable order contribution of the metric in Eq.~\eqref{FGexpansion} when applied to Eq.~\eqref{weyleinsteincondition} through Eq.~\eqref{Hessianzz}. Furthermore, the radial equation admits no independent logarithmic partners of the $z^{\pm1}$ branches within the constrained asymptotic sector.

The mixed and tangential components of Eq.~\eqref{weyleinsteincondition} do not generate additional non-normalizable radial branches; instead, they impose additional compatibility conditions involving the FG coefficients of Eq.~\eqref{FGexpansion} and the radial expansion of $\Sigma$ in Eq.~\eqref{Sigmaexp}.

In particular, for the $zi$ component we obtain
\begin{equation}
\partial_z \partial_j \Sigma -\Gamma_{zj}^m \partial_m \Sigma \simeq 0 \,.
\end{equation}
Substituting Eqs.~\eqref{FGexpansion} and~\eqref{Sigmaexp} into the last equation gives at leading order
\begin{equation}
2 \mathcal{D}_{\left(0\right)j} \sigma_{\left(2\right)}- \frac{1}{\ell^2} g_{\left(2\right)j}^m \mathcal{D}_{\left(0\right)m} \sigma_{\left(0\right)} =0 \,.
\end{equation}
Similarly, the $ij$ component reads
\begin{equation}
\partial_i \partial_j \Sigma -\Gamma^{z}_{ij}\partial_z \Sigma- \Gamma^{m}_{ij}\partial_m \Sigma-\frac{\Sigma}{\ell^2}h_{ij} \simeq 0 \,.
\label{ijcomponent}
\end{equation}
Therefore, the leading order contribution yields
\begin{equation}
\mathcal{D}_{\left(0\right)i} \mathcal{D}_{\left(0\right)j} \sigma_{\left(0\right)}-\frac{1}{\ell^2} g_{\left(2\right)ij} \sigma_{\left(0\right)}-2g_{\left(0\right)ij} \sigma_{\left(2\right)} =0\,.
\label{ijeom}
\end{equation}
It follows, by taking the trace, that $\sigma_{\left(2\right)}$ is not an independent coefficient in the expansion but is fixed by $\sigma_{\left(0\right)}$ according to
\begin{equation}
\sigma_{\left(2\right)}=\frac{1}{2d} \left(\Box_{\left(0\right)}+\mathcal{S}_{\left(0\right)}\right)\sigma_{\left(0\right)} \,,
\end{equation}
where $\mathcal{S}_{\left(0\right)} = g_{\left(0\right)}^{ij} \mathcal{S}_{\left(0\right)ij}$.
Substituting this relation into Eq.~\eqref{ijeom} gives
\begin{equation}
\left[\mathcal{D}_{\left(0\right)i} \mathcal{D}_{\left(0\right)j}\sigma_{\left(0\right)}+\mathcal{S}_{\left(0\right)ij} \sigma_{\left(0\right)}\right]^{\left(tf\right)} =0 \,,
\label{tracefreesigma0}
\end{equation}
where $\left(tf\right)$ denotes the trace-free part of the corresponding tensor. This relation provides a necessary compatibility condition on the leading boundary Weyl parameter. Higher FG orders may impose further compatibility conditions on the asymptotic extension of $\Sigma$.

Eq.~\eqref{tracefreesigma0} is the almost-Einstein scale equation for the boundary conformal structure~\cite{Gover:2008nsa}. A restricted Weyl parameter $\Sigma$ with nonzero leading boundary data $\sigma_{(0)}$ therefore requires the boundary conformal class to admit a nontrivial almost-Einstein scale. On any open set where such a scale $\sigma_{(0)}$ is nonvanishing, the metric $\widehat g_{\left(0\right)ij}=\sigma_{(0)}^{-2}g_{\left(0\right)ij}$ is an Einstein representative of the boundary conformal class. This representative may have nonzero Weyl curvature, so the boundary condition also accommodates non-conformally-flat geometries.

A complementary geometric interpretation of the restricted Weyl condition follows from its relation to diffeomorphisms. Introducing the gradient vector field $\xi^\mu=\ell^2\nabla^\mu\Sigma$, Eq.~\eqref{weyleinsteincondition} implies
\begin{equation}
\delta_\Sigma g_{\mu\nu}-\mathcal L_\xi g_{\mu\nu}
=-2\ell^2\left(\nabla_\mu\nabla_\nu\Sigma-\ell^{-2}g_{\mu\nu}\Sigma\right)\simeq0 \,.
\label{diffweyl}
\end{equation}
Eq.~\eqref{diffweyl} provides a geometric interpretation of the restricted Weyl transformation as an asymptotic gradient diffeomorphism. Using the expansion ~\eqref{Sigmaexp}, the corresponding vector field has components $\xi^z=-\sigma_{\left(0\right)}+\mathcal{O}\left(z^2\right)$ and $\xi^i=z\mathcal{D}_{\left(0\right)}^i\sigma_{\left(0\right)}+\mathcal{O} \left(z^3\right)$. Thus, its leading radial component describes a radial displacement near the conformal boundary, accompanied by a subleading tangential shift.

Eq.~\eqref{diffweyl} further indicates that the transformed metric, after the corresponding infinitesimal coordinate adaptation, preserves the asymptotic Einstein--AdS relations. For example, the second-order coefficient obeys $\widetilde g_{\left(2\right)ij}=-\ell^2 \mathcal{S}_{ij}[\widetilde g_{\left(0\right)}]$, without requiring $\widetilde g_{\left(0\right)ij}=g_{\left(0\right)ij}$. The same requirement extends to all non-normalizable orders, while the independent normalizable data are subject to the compatibility conditions required for the existence of an admissible restricted Weyl parameter~\eqref{weyleinsteincondition}.

This equivalence holds at the infinitesimal level and does not require the preservation of a fixed FG defining function or the existence of finite Weyl transformations. It also differs from the Penrose--Brown--Henneaux (PBH) transformations~\cite{Imbimbo:1999bj}, which preserve the FG gauge and whose radial generators vanish at the conformal boundary. By contrast, the gradient vector field associated with the Weyl parameter $\Sigma$ generally has a nonvanishing leading radial component. The resulting relation therefore characterizes the restricted conformal transformations compatible with the asymptotic Einstein--AdS structure, without introducing an independent Weyl gauge symmetry of Einstein gravity.

\section{Topological renormalization and the restricted Weyl mismatch}
\label{TopologicalRenormalization}

We now apply the restricted Weyl completion scheme introduced in Sec.~\ref{weyleinsteinorbit}, in $D=2n$ dimensions. As a first step, we consider only the Euler part of the $\mathcal{L}_{weyl}$ with an undetermined coupling $c_{2n}$, as
\begin{equation}
\mathcal{L}_{weyl}=c_{2n} \mathcal{E}_{2 n} \,, \quad \mathcal{E}_{2 n} = \frac{1}{2^n} \delta _{\nu _{1} \ldots  \nu _{2 n}}^{\mu _{1} \ldots  \mu _{2 n}} R_{\mu _{1} \mu _{2}}^{\nu _{1} \nu _{2}} \cdots  R_{\mu _{2 n -1} \mu _{2 n}}^{\nu _{2 n -1} \nu _{2 n}}  \,.
\end{equation}
We fix $c_{2n}$ by the completion criterion~\eqref{variationdiv} on the corresponding density
\begin{equation}
\mathcal{L}^{\left(2n\right)}_{\text{top}} =\sqrt{\vert g \vert} \left (R -2 \Lambda  +c_{2 n} \mathcal{E}_{2 n}\right )\,.
\label{L2ntop}
\end{equation}
To do so, we first consider an arbitrary infinitesimal Weyl rescaling and then impose the asymptotic condition derived in Sec.~\ref{weyleinsteinorbit}.

In particular, using the variation of the Ricci scalar
\begin{equation}
\delta _{\sigma } R= -2 \sigma R - 2\left(D-1\right) \Box \sigma \,,
\label{weylvarRicci}
\end{equation}
the Einstein--Hilbert density with negative cosmological constant varies as
\begin{equation}
\delta _{\sigma } \left [\sqrt{\vert g \vert} (R -2 \Lambda )\right ] 
=2 \sqrt{\vert g \vert} \left [\left (n -1\right ) \sigma  \left (R +\dfrac{2 n \left (2 n -1\right )}{\ell ^{2}}\right ) -\left (2 n -1\right )  \Box\sigma \right ] \,.
\label{LEH2nD}
\end{equation}
Likewise, the Euler density transforms according to
\begin{equation}
\delta _{\sigma } \left(\sqrt{\vert g \vert} \mathcal{E}_{2 n} \right) = -\dfrac{n \sqrt{\vert g \vert}}{2^{n -2}} \delta _{\nu _{1} \ldots  \nu _{2 n -1}}^{\mu _{1} \ldots  \mu _{2 n -1}}  \nabla _{\mu _{2 n -1}}\left (R_{\mu _{1} \mu _{2}}^{\nu _{1} \nu _{2}} \cdots  R_{\mu _{2 n -3} \mu _{2 n -2}}^{\nu _{2 n -3} \nu _{2 n -2}}  \nabla ^{\nu _{2 n -1}}\sigma \right )\,.
\label{varEulerevenD}
\end{equation}
On a closed manifold the integrated Euler density is conformally invariant, while in the presence of a boundary its variation reduces to a boundary term. More details on the Weyl rescaling of the Euler density can be found in Appendix~\ref{Eulerweylvariation}.

Combining Eqs.~\eqref{LEH2nD} and~\eqref{varEulerevenD} shows that the full density is not invariant under an arbitrary local rescaling,
\begin{equation}
\delta _{\sigma } \mathcal{L}^{\left(2n\right)}_{\text{top}} \neq 0.
\end{equation}
We now restrict $\sigma \rightarrow \Sigma$ and use the asymptotic identities of Section~\ref{weyleinsteinorbit}. Applying the differential Bianchi identity for the Riemann tensor to Eq.~\eqref{varEulerevenD} gives
\begin{equation}
\delta _{\Sigma } \left(\sqrt{\vert g \vert} \mathcal{E}_{2 n} \right)=-\dfrac{n \sqrt{\vert g \vert}}{2^{n -2}} \delta _{\nu _{1} \ldots  \nu _{2 n -1}}^{\mu _{1} \ldots  \mu _{2 n -1}}   R_{\mu _{1} \mu _{2}}^{\nu _{1} \nu _{2}} \cdots  R_{\mu _{2 n -3} \mu _{2 n -2}}^{\nu _{2 n -3} \nu _{2 n -2}}  \nabla _{\mu _{2 n -1}} \nabla ^{\nu _{2 n -1}}\Sigma \,.
\label{varEulerevenD2}
\end{equation}
The double covariant derivative is further simplified using the condition~\eqref{weyleinsteincondition}. Finally, reexpressing the Riemann tensor in terms of the AdS curvature $F^{\alpha \beta}_{\mu \nu}$ yields
\begin{equation}
\delta _{\Sigma } \left(\sqrt{\vert g \vert } \mathcal{E}_{2 n} \right) \simeq\Sigma \sqrt{\vert g \vert} \sum \limits _{i =0}^{n -1} \dfrac{( -1)^{n -i} n ! (2 n -2 i) !}{i ! (n -i -1) ! 2^{i -1} \ell ^{2 (n -i)}} \delta _{\nu _{1} \ldots  \nu _{2 i}}^{\mu _{1} \ldots  \mu _{2 i}}  F_{\mu _{1} \mu _{2}}^{\nu _{1} \nu _{2}} \cdots  F_{\mu _{2 i -1} \mu _{2 i}}^{\nu _{2 i -1} \nu _{2 i}} \,.
\end{equation}
Evaluating the curvature tensors on an Einstein–AdS representative and applying Eq.~\eqref{weyleinsteincondition} to Eq.~\eqref{LEH2nD}, we obtain
\begin{equation}
\delta _{\Sigma } \left. \mathcal{L}^{\left(2n\right)}_{\text{top}} \right \vert_{E} \simeq \Sigma \sqrt{\vert g \vert} \left\{\frac{4n (2 n -1)}{\ell^2} \left[ \frac{( -1)^{n} (2 n -2) ! n c_{2 n} }{\ell ^{2 (n -1)}}-1\right]+ c_{2 n} \sum \limits _{i =2}^{n -1} \dfrac{( -1)^{n -i} n ! (2 n -2 i) !}{i ! (n -i -1) ! 2^{i -1} \ell ^{2 (n -i)}}\Phi _{\left (i\right )} \right\}\,.
\label{Ltop1}
\end{equation}
where the expression~\eqref{phiF} has been taken into account. The sum starts from $i=2$ due to the vanishing of the single trace of the AdS curvature tensor since $F^{\alpha \beta}_{\mu \nu}\vert_E= W^{\alpha \beta}_{\mu \nu} $.
Here, $\simeq$ denotes the asymptotic equality inherited from Eq.~\eqref{weyleinsteincondition}. At this intermediate stage, the omitted density terms, associated with the curvature-independent part,
may include contributions of order $\mathcal{O}(z^{-1})$. However, when the Euler coupling is fixed by
\begin{equation}
c_{2n} = \frac{\left(-1\right)^{n} \ell^{2n-2}}{\left(2n-2\right)!n} \,,
\label{c2n}
\end{equation}
these terms vanish while the curvature-dependent remainders are asymptotically integrable. Thus, for this value of $c_{2n}$, the Einstein--Hilbert density supplemented by
the Euler invariant can be written as
\begin{equation}
\mathcal L_{\mathrm{top}}^{(2n)}
=\sqrt{|g|}\sum_{i=2}^{n}\beta_i\Phi_{\left(i\right)},\qquad
 \beta_i=
 \frac{(-1)^i(n-1)!(2n-2i)!\,\ell^{2i-2}}
 {2^i i!(n-i)!(2n-2)!} \,.
 \label{top-polynomial}
\end{equation}
Hence, the variation along $\Sigma$ reduces to
\begin{align}
\left. \delta _{\Sigma } \mathcal{L}^{\left(2n\right)}_{\text{top}} \right \vert_{E} &\doteq 2\sqrt{|g|}\sum_{i=2}^{n}(n-i)\beta_i\,\Sigma\Phi_{\left(i\right)} \label{weylvariationIktphi}\\
&=\sqrt{\vert g \vert} \frac{\left (n -1\right )!}{\left (2n -2\right ) !}\sum \limits _{i =2}^{n -1}\frac{\left ( -1\right )^{i} \left [2\left (n -i\right )\right ] !\ell ^{2\left (i -1\right )}}{i !\left (n -i -1\right ) !2^{i -1}}\Sigma \Phi _{\left (i\right )} \,.
\label{weylvariationIkt}
\end{align}
The restricted Weyl completion fixes the Euler-density coupling precisely to the value~\eqref{c2n} associated with topological renormalization~\cite{Olea:2005gb}. In four bulk dimensions, the curvature-dependent sum is empty, so the topologically renormalized density
\begin{equation}
\mathcal{L}^{\left(4\right)}_{\text{top}} =\sqrt{\vert g \vert} \left (R -2 \Lambda  +\frac{\ell^2}{4} \mathcal{E}_{4}\right )\,,
\label{gbrenormaliza}
\end{equation}
already satisfies the condition~\eqref{variationdiv}. This prescription yields finite conserved charges and a finite Euclidean on-shell action~\cite{Aros:1999id,Aros:1999kt}. The action admits a Dirichlet variational principle for $g_{\left(0\right)ij}$~\cite{Anastasiou:2017rjf}. It also agrees with standard holographic renormalization at the conformal boundary~\cite{Miskovic:2009bm}.

In higher even dimensions, the obstruction of  Eq.~\eqref{weylvariationIkt} is non-vanishing. To determine whether these obstructions are asymptotically integrable, by contributing at finite order, we examine their radial falloff. Since the AdS curvature coincides with the Weyl tensor, on EAdS configurations, the odd-power coefficients of the FG expansion~\eqref{FGexpansion} below the normalizable order vanish. Therefore, the leading order falloff for the three independent components of the Weyl tensor reads
\begin{align}
W^{iz}_{jz}|_{E}=F^{iz}_{jz}
&= \frac{z^4}{(d-4)\ell^2}\mathcal{B}^{i}_{(0)j}
+\mathcal{O}(z^6)\,, \notag\\
W^{ij}_{km}|_{E}=F^{ij}_{km}
&= \frac{z^2}{\ell^2}\mathcal{W}^{ij}_{(0)km}
+\mathcal{O}(z^4)\,, \notag\\
W^{iz}_{jk}|_{E}=F^{iz}_{jk}
&= \frac{z^3}{\ell^2}\mathcal{C}^{i}_{(0)jk}
+\mathcal{O}(z^5)\,.
\label{weylcomponentseinstein}
\end{align}
These relations are determined by substituting $g_{\left(1\right)ij}=0$ into the independent components of the Weyl tensor in the Appendix~\ref{GC}. For AlAdS boundary geometries, the leading tangential components behave as $\mathcal{O}(z^2)$, so the curvature polynomials $\Phi_{\left(i\right)}$ generally contain contributions of order $\mathcal{O}(z^{2i})$. Since the volume density scales as $\sqrt{g}\sim z^{-D}$ and the scaling parameter behaves as $\Sigma\sim z^{-1}$, these terms contribute at orders capable of generating IR divergences. Hence, the scalar $\Phi_{\left(i\right)}$  in Eq.~\eqref{weylvariationIkt} obstruct restricted Weyl equivalence~\eqref{variationdiv} of the topologically renormalized density.

For ACF geometries, by contrast, the vanishing of the boundary Weyl tensor implies that the bulk Weyl tensor falls off at normalizable order~\cite{Anastasiou:2020zwc}. Indeed, the RHS of Eq.~\eqref{weylvariationIkt} behaves as $\mathcal{O} (z^{d(i-1)-2})$. Namely, the obstruction terms are sufficiently suppressed to contribute asymptotically only at finite order or beyond, such that
\begin{equation}
\left. \delta _{\Sigma } \mathcal{L}^{\left(2n\right)}_{\text{top}} \right \vert_{E}\doteq0 \,.
\end{equation}
This is consistent with the established sufficiency of topological renormalization for ACF geometries~\cite{Anastasiou:2020zwc}. The recovery of the curvature-dependent part of topological renormalization in the ACF sector naturally motivates adopting its associated constant contribution. We therefore fix the otherwise undetermined additive constant $I_0^{(2n)}$ in Eq.~\eqref{I2naction} to the value prescribed by that prescription as,
\begin{equation}
I_0^{(2n)}=-\frac{c_{2n}}{16\pi G_N}(4\pi)^n n!
\chi(\mathcal{M}_{2n}) \,.
\label{I02n}
\end{equation}
Since the Euler characteristic is invariant under smooth metric deformations at fixed topology, $I_0^{(2n)}$ does not contribute to the restricted Weyl variation of the action. In particular, it plays no role in determining the completion coefficients or in cancelling infrared divergences. Its inclusion fixes the finite normalization of the resulting renormalized action consistently with the topological renormalization prescription and facilitates comparison with standard holographic renormalization. This choice is also relevant for black-hole thermodynamics and holographic quantum information observables~\cite{Anastasiou:2021tlv,Anastasiou:2023oro,anastasiou2018topological,Anastasiou:2024rxe}.

In the following sections, we construct covariant total derivatives that compensate the obstruction~\eqref{weylvariationIkt} in every even bulk dimension. Their renormalization properties are then tested through comparisons with known actions and explicit black-hole applications. The resulting gravitational completion takes the form
\begin{equation}
\mathcal{L}_{\mathrm{weyl}}
=c_{2n}\mathcal{E}_{2n}+\nabla_{\mu}J^{\mu}\,,
\end{equation}
where $c_{2n}$ is fixed by Eq.~\eqref{c2n} and $J^\mu$ denotes the covariant current encoding the required total-derivative counterterms.

\section{Restricted Weyl completion in six dimensions}
\label{6D}

The first nontrivial realization of this prescription is performed in six dimensions. Setting $n=3$, Eq.~\eqref{weylvariationIkt} reduces to
\begin{equation}
\delta _{\Sigma } \left. \mathcal{L}^{\left(6\right)}_{\text{top}} \right \vert _{E} \doteq \sqrt{\vert g \vert} \frac{\ell ^{2}}{24}\Sigma \Phi _{\left (2\right )} \,,
\label{weylvarIktpartial1}
\end{equation}
where the obstruction is proportional to the quadratic curvature invariant $\Phi_{(2)}$. We therefore seek a covariant total derivative whose restricted Weyl variation generates this structure. The simplest candidate is the Laplacian descendant $\Box\Phi_{\left(2\right)}$. The choice of this candidate is also natural from the structure of the Yamabe operator
\begin{equation}
P_{2} = - \Box+\frac{D -2}{4\left(D-1\right)}R \,,
\end{equation}
which is a conformally covariant differential operator~\cite{Gover:2002ay}. Indeed, for a scalar field of scaling weight $\Delta$ ($\tilde{\Phi}=e^{\Delta \sigma} \Phi$), the infinitesimal Weyl variation of the Laplacian reads
\begin{equation}
\delta _{\sigma }\left (\Box \Phi \right ) =\left [\left (\Delta  -2\right )\sigma  \Box\Phi  +\Delta \Phi  \Box \sigma  +\left (D +2\Delta  -2\right ) \nabla ^{\lambda }\sigma  \nabla _{\lambda }\Phi \right ] \,. \label{weylvarboxPhi}
\end{equation}
The key feature is that for asymptotically EAdS configurations, the scalar $\Phi _{\left (i\right )}$ has a Weyl weight $\Delta_{\left(i\right)}=-2i$ under the transformation~\eqref{weylvarF2}, such that
\begin{equation}
\delta_{\Sigma} \Phi _{\left (i\right )} \simeq -2i\Sigma \Phi _{\left (i\right )} \,.
\label{Phiidimension}
\end{equation}
For $\Phi_{\left(2\right)}$, the weight is $\Delta_{\left(2\right)}=-4$, such that the homogeneous term in Eq.~\eqref{weylvarboxPhi} is actually $-6\Sigma\Box\Phi_{\left(2\right)}$. In six dimensions, this term is cancelled by the variation of the volume
\begin{equation}
\delta_\Sigma\sqrt{|g|}=6\Sigma\sqrt{|g|}\,.
\end{equation}
As a consequence, the corresponding scalar density transforms as
\begin{equation}
\delta _{\Sigma }\left (\sqrt{\vert g \vert} \Box\Phi _{\left (2\right )}\right ) \doteq  -4\sqrt{\vert g \vert}\left (\Phi _{\left (2\right )} \Box\Sigma  + \nabla ^{\lambda }\Sigma  \nabla _{\lambda }\Phi _{\left (2\right )}\right ) \,.
\label{var6dpartial}
\end{equation}
Using Eq.~\eqref{weyleinsteincondition}, this can be cast in the form
\begin{equation}
\delta _{\Sigma }\left (\sqrt{\vert g \vert} \Box\Phi _{\left (2\right )}\right ) \doteq -4\sqrt{\vert g \vert}\left (\frac{6 \Sigma}{\ell^2}\Phi _{\left (2\right )}   + \nabla ^{\lambda }\Sigma  \nabla _{\lambda }\Phi _{\left (2\right )}\right ) \,. \label{var6dpartial1}
\end{equation}
This expression simplifies further by exploiting the antisymmetric structure of the scalar $\Phi _{\left (i\right )}$ in Eq.~\eqref{phiF}. In particular, the gradient of $\Phi _{\left (2\right )}$ is
\begin{equation}
\nabla _{\lambda }\Phi _{\left (2\right )} =\delta _{\mu _{1}\ldots \mu _{4}}^{\nu _{1}\ldots \nu _{4}} \nabla _{\lambda }\left (F_{\nu _{1}\nu _{2}}^{\mu _{1}\mu _{2}}F_{\nu _{3}\nu _{4}}^{\mu _{3}\mu _{4}}\right ) =2\delta _{\mu _{1}\ldots \mu _{4}}^{\nu _{1}\ldots \nu _{4}}F_{\nu _{1}\nu _{2}}^{\mu _{1}\mu _{2}} \nabla _{\lambda }F_{\nu _{3}\nu _{4}}^{\mu _{3}\mu _{4}} \,.
\label{bianchi}
\end{equation}
The Bianchi identity gives
\begin{align} 
&\nabla _{\lambda }F_{\nu _{3}\nu _{4}}^{\mu _{3}\mu _{4}} + \nabla _{\nu _{3}}F_{\nu _{4}\lambda }^{\mu _{3}\mu _{4}} + \nabla _{\nu _{4}}F_{\lambda \mu _{3}}^{\mu _{3}\mu _{4}} =0 \Rightarrow  \nonumber  \\
&\Rightarrow  \nabla _{\lambda }F_{\nu _{3}\nu _{4}}^{\mu _{3}\mu _{4}} =2 \nabla _{[\nu _{4}}F_{\nu _{3}]\lambda }^{\mu _{3}\mu _{4}} \,.
\label{bianchi2}
\end{align}
Combining the last two equations, Eq.~\eqref{bianchi} can be written as
\begin{equation}
\nabla _{\lambda }\Phi _{\left (2\right )} =2\delta _{\mu _{1}\ldots \mu _{4}}^{\nu _{1}\ldots \nu _{4}}F_{\nu _{1}\nu _{2}}^{\mu _{1}\mu _{2}} \nabla _{\lambda }F_{\nu _{3}\nu _{4}}^{\mu _{3}\mu _{4}} =4\delta _{\mu _{1}\ldots \mu _{4}}^{\nu _{1}\ldots \nu _{4}}F_{\nu _{1}\nu _{2}}^{\mu _{1}\mu _{2}} \nabla _{\nu _{4}}F_{\nu _{3}\lambda }^{\mu _{3}\mu _{4}} \,.
\end{equation}
Therefore the derivative contribution of Eq.~\eqref{var6dpartial1} becomes
\begin{gather}
\nabla _{\lambda }\Phi _{\left (2\right )} \nabla ^{\lambda }\Sigma 
=4\delta _{\mu _{1}\ldots \mu _{4}}^{\nu _{1}\ldots \nu _{4}} \nabla _{\nu _{4}}\left (F_{\nu _{1}\nu _{2}}^{\mu _{1}\mu _{2}}F_{\nu _{3}\lambda }^{\mu _{3}\mu _{4}}\right ) \nabla ^{\lambda }\Sigma  \nonumber  \\
=4\delta _{\mu _{1}\ldots \mu _{4}}^{\nu _{1}\ldots \nu _{4}}\left [ \nabla _{\nu _{4}}\left (F_{\nu _{1}\nu _{2}}^{\mu _{1}\mu _{2}}F_{\nu _{3}\lambda }^{\mu _{3}\mu _{4}} \nabla ^{\lambda }\Sigma \right ) -F_{\nu _{1}\nu _{2}}^{\mu _{1}\mu _{2}}F_{\nu _{3}\lambda }^{\mu _{3}\mu _{4}} \nabla _{\nu _{4}} \nabla ^{\lambda }\Sigma \right ] \,.
\label{delphi2delsigmainterm}
\end{gather}
The asymptotic behavior of $F^{\alpha \beta}_{\mu \nu}$ in Eq.~\eqref{weylcomponentseinstein}, together with the identity~\eqref{FcottonEinstein}, implies that the total-derivative contribution in Eq.~\eqref{delphi2delsigmainterm} is of order $\mathcal O(z^6)$. Since the six-dimensional volume density scales as $\sqrt{\vert g \vert}=\mathcal O(z^{-6})$, the corresponding contribution to the variation of the gravitational density is at most of order $\mathcal O(z^0)$ and is therefore integrable at the conformal boundary. Therefore, it can be omitted when determining the completion coefficient. Finally, substituting Eq.~\eqref{weyleinsteincondition} into~\eqref{delphi2delsigmainterm} leads directly to
\begin{equation}
\nabla _{\lambda }\Phi _{\left (2\right )} \nabla ^{\lambda }\Sigma \simeq -\frac{4\Sigma }{\ell ^{2}}\Phi _{\left (2\right )} \,.
\label{delphi2delsigma}
\end{equation}
Substituting this relation into Eq.~\eqref{var6dpartial1} gives
\begin{equation}
\delta _{\Sigma }\left (\sqrt{\vert g \vert} \Box\Phi _{\left (2\right )}\right )\doteq -\frac{8\sqrt{\vert g \vert}}{\ell ^{2}}\Sigma \Phi _{\left (2\right )} \,.
\end{equation}
The restricted Weyl variation of the total-derivative term cancels the mismatch~\eqref{weylvarIktpartial1}. The completed density therefore satisfies
\begin{equation}
\left.
\delta_\Sigma\left[
\mathcal{L}^{(6)}_{\mathrm{top}}
+\frac{\ell^4}{192}\sqrt{|g|}\,\Box\Phi_{\left(2\right)}
\right]
\right|_E
\doteq 0\,,
\label{L6completed}
\end{equation}
in agreemet with Eq.~\eqref{variationdiv}. Integrating the Weyl-completed density in brackets in Eq.~\eqref{L6completed} and including the constant~\eqref{I02n} gives
\begin{equation}
I_{\mathrm{ren}}^{(6)}
=
\frac{1}{16\pi G_N}
\int_{\mathcal{M}_6} d^6X\,\sqrt{|g|}
\left[
R-2\Lambda+c_6\mathcal{E}_6
+\frac{\ell^4}{48}\Box\left(
F_{\mu\nu}^{\alpha\beta}F_{\alpha\beta}^{\mu\nu}
\right)
\right]
+\frac{\pi^2\ell^4}{3G_N}\chi(\mathcal{M}_6)\,.
\label{weylL6}
\end{equation}
The resutling action coincides with the divergence-free action obtained though Conformal Renormalization~\cite{Anastasiou:2020mik}, providing the first nontrivial test of the derivative correction fixed by the restricted Weyl criterion.

The only remaining ambiguity would be the addition of an independent conformally invariant total derivative, since a genuine bulk term would modify the EOM. However, no such independent contribution exists within the 6D basis of conformal invariants~\cite{Graham1985,parker1987invariants,Erdmenger:1997gy}. This conclusion is also consistent with the CR prescription. The six-dimensional parent theory, the Lü--Pang--Pope (LPP) conformal gravity~\cite{Lu:2013hx}, admits an additional boundary term involving the Cotton tensor~\cite{Anastasiou:2020mik}, but this contribution vanishes identically on EAdS manifolds. Furthermore, Boulanger and Rovere showed in Ref.~\cite{Boulanger:2025oli} that LPP conformal gravity is the unique 6D conformally invariant combination admitting an EAdS sector. The ambiguity in Eq.~\eqref{weylL6} is therefore fixed.

In addition to its IR finiteness~\cite{Anastasiou:2025usa}, the six-dimensional action~\eqref{weylL6} admits well-defined Dirichlet boundary conditions for the conformal boundary metric $g_{\left(0\right)ij}$. To verify this property, we examine the asymptotic variations of the bulk and total-derivative contributions separately. 

Indeed, the bulk contribution behaves asymptotically as
\begin{align}\label{VarBulk}
    \delta I^{\rm bulk} &= - \frac{\ell^4}{384 \pi G_N}  \int\limits_{\partial \mathcal{M}_6}d^5 x\sqrt{
    \vert g_{\left(0\right)} \vert}\left[\frac{1}{z}\bigg(\frac{1}{4}\delta^{i_1 ...i_5}_{j_1 ...j_5}\mathcal{W}^{j_1 j_2}_{\left(0\right) i_1 i_2}\mathcal{W}^{j_3 j_4}_{\left(0\right) i_3 i_4}(g^{-1}_{\left(0\right)}\delta g_{\left(0\right)})^{j_5}_{i_5} \right. \notag \\
    &\left.- 8\left(\mathcal{B}^{i}_{\left(0\right)j}+\mathcal{D}^{m}_{\left(0\right)}\mathcal{C}^{i}_{\left(0\right)jm}\right)(g^{-1}_{\left(0\right)}\delta g_{\left(0\right)})^{j}_{i}\bigg)+\frac{60}{\ell^5}g^i_{\left(5\right)j}\left(g^{-1}_{\left(0\right)}\delta g_{\left(0\right)}\right)^{j}_{i}+\mathcal{O}\left(z\right) \right] \,.
\end{align}
Similarly, the variation of the total derivative term gives
\begin{align}\label{VarBdry}
    \delta I^{\rm bdry} = \frac{\ell^4}{384 \pi G_N}  \int\limits_{\partial \mathcal{M}_6}d^5 x\sqrt{\vert g_{\left(0\right)} \vert}\Bigg[\frac{1}{z}\bigg(\frac{1}{4}\delta^{i_1 ...i_5}_{j_1 ...j_5}\mathcal{W}^{j_1 j_2}_{\left(0\right)i_1 i_2}\mathcal{W}^{j_3 j_4}_{\left(0\right) i_3 i_4}(g^{-1}_{\left(0\right)}\delta g_{\left(0\right)})^{j_5}_{i_5} \notag \\ - 8\left(\mathcal{B}^{i}_{\left(0\right)j}+\mathcal{D}^{m}_{\left(0\right)}\mathcal{C}^{i}_{\left(0\right)jm}\right)(g^{-1}_{\left(0\right)}\delta g_{\left(0\right)})^{j}_{i}\bigg) + \mathcal{O}\left(z\right) \Bigg] \,.
\end{align}
Their divergent terms cancel, leaving the finite expression for the holographic stress tensor as
\begin{align}\label{Thol}
     \delta I_{\rm ren}^{\left(6\right)}  = \frac{1}{2}\int\limits_{\partial \mathcal{M}_6}d{^5x}\sqrt{\vert g_{\left(0\right)}\vert}\delta g_{\left(0\right)ij}\langle T^{ij} \rangle\,, \;\;\;\;\; \mbox{where} \;\;\;\;\; \langle T_{ij} \rangle = -\frac{5}{16\pi G_N\,\ell} \, g_{\left(5\right)ij} \,,
\end{align}
in agreement with HR. The total derivative fixed by restricted Weyl completion cancels the divergent terms in the Dirichlet variation and yields the standard holographic stress tensor determined by $g_{(5)ij}$. Together with the agreement of Eq.~\eqref{weylL6} with the conformally renormalized action, this establishes both on-shell finiteness and a finite Dirichlet variation for general boundary geometries.

\section{Restricted Weyl completion in arbitrary even dimensions}
\label{evenD}

\subsection{Recursive closure and factorization}
\label{closure}

Having established the $D=4$ and $D=6$ cases, we now extend the new procedure to higher even bulk dimensions. The crucial step is the derivation of recursive identities for the Laplacian descendants of the curvature polynomials $\Phi_{\left(i\right)}$. Solving these identities, we construct combinations with definite restricted Weyl weights that allow us to generate expressions applicable to arbitrary even dimensions. 
The starting point is the generalization of the gradient identity in Eq.~\eqref{delphi2delsigma}.
Applying the gradient to $\Phi_{\left(i\right)}$, we obtain
\begin{equation}
\nabla _{\lambda }\Phi _{\left (i\right )} = i\delta _{\mu _{1}\ldots \mu _{2i}}^{\nu _{1}\ldots \nu _{2i}}F_{\nu _{1}\nu _{2}}^{\mu _{1}\mu _{2}} \cdots F_{\nu _{2i -3}\nu _{2i -2}}^{\mu _{2i -3}\mu _{2i -2}} \nabla _{\lambda }F_{\nu _{2i -1}\nu _{2i}}^{\mu _{2i -1}\mu _{2i}} \,,
\end{equation}
which, using the Bianchi identity~\eqref{bianchi2}, can be rewritten as
\begin{equation}
\nabla _{\lambda }\Phi _{\left (i\right )}
=2i\delta _{\mu _{1}\ldots \mu _{2i}}^{\nu _{1}\ldots \nu _{2i}} \nabla _{\nu _{2i}}\left (F_{\nu _{1}\nu _{2}}^{\mu _{1}\mu _{2}} \cdots F_{\nu _{2i -3}\nu _{2i -2}}^{\mu _{2i -3}\mu _{2i -2}}F_{\nu _{2i -1}\lambda }^{\mu _{2i -1}\mu _{2i}}\right ) \,.
\end{equation}
As in six dimensions, contracting this expression with the gradient of $\Sigma$ gives
\begin{align}
\nabla ^{\lambda }\Sigma  \nabla _{\lambda }\Phi _{\left (i\right )}
&=2i\delta _{\mu _{1}\ldots \mu _{2i}}^{\nu _{1}\ldots \nu _{2i}}\left[ \nabla _{\nu _{2i}}\left (F_{\nu _{1}\nu _{2}}^{\mu _{1}\mu _{2}} \cdots F_{\nu _{2i -3}\nu _{2i -2}}^{\mu _{2i -3}\mu _{2i -2}}F_{\nu _{2i -1}\lambda }^{\mu _{2i -1}\mu _{2i}} \nabla ^{\lambda }\Sigma \right ) \right.\nonumber  \\
&\left. -F_{\nu _{1}\nu _{2}}^{\mu _{1}\mu _{2}} \cdots F_{\nu _{2i -3}\nu _{2i -2}}^{\mu _{2i -3}\mu _{2i -2}}F_{\nu _{2i -1}\lambda }^{\mu _{2i -1}\mu _{2i}} \nabla _{\nu _{2i}} \nabla ^{\lambda }\Sigma \right] \,.
\label{delsigmadelphieq}
\end{align}
The total derivative contribution in the last equation is controlled by the condition~\eqref{FcottonEinstein}. Using Eq.~\eqref{weylcomponentseinstein}, we find that this term falls off as $\mathcal O(z^{D+2(i-2)})$. After multiplication by the bulk volume density, is scales as $\mathcal O(z^{2(i-2)})$, which is integrable at the conformal boundary for every $i\geq2$. This term therefore belongs to the remainder admitted by the restricted Weyl equivalence criterion~\eqref{variationdiv} and does not contribute to the determination of the Weyl completion coefficients. Hence, the remaining term gives
\begin{equation}
\nabla ^{\lambda }\Sigma  \nabla _{\lambda }\Phi _{\left (i\right )} \simeq -2i\delta _{\mu _{1}\ldots \mu _{2i}}^{\nu _{1}\ldots \nu _{2i}}F_{\nu _{1}\nu _{2}}^{\mu _{1}\mu _{2}} \cdots F_{\nu _{2i -3}\nu _{2i -2}}^{\mu _{2i -3}\mu _{2i -2}}F_{\nu _{2i -1}\lambda }^{\mu _{2i -1}\mu _{2i}} \nabla _{\nu _{2i}} \nabla ^{\lambda }\Sigma  \,.
\end{equation}
Taking into account Eq.~\eqref{weyleinsteincondition}, we obtain
\begin{equation}
\nabla ^{\lambda }\Sigma  \nabla _{\lambda }\Phi _{\left (i\right )} \simeq \frac{\Delta_{\left(i\right)}\Sigma }{\ell ^{2}}\Phi _{\left (i\right )} \,.
\label{delsigmadelphi}
\end{equation}
To simplify the resulting expressions in arbitrary dimension, we define
\begin{equation}
O_{\left (i\right )}^{q}=\Box^{q} \Phi _{\left (i\right )} \,, \quad K_{\left (i\right )}^{q}=\nabla^\lambda \Sigma\nabla_\lambda O_{\left (i\right )}^{q} \,.
\label{Oqidef}
\end{equation}
In this notation, Eq.~\eqref{delsigmadelphi} can be equivalently written as
\begin{equation}
K_{\left (i\right )}^{0} \simeq \frac{\Delta_{\left(i\right)}\Sigma }{\ell ^{2}}\Phi _{\left (i\right )} \,.
\label{K0}
\end{equation}
The higher-dimensional completion is governed by three recursive identities that determine the restricted Weyl variations of successive Laplacians. The first relation reads
\begin{equation}
\delta_\Sigma O_{\left (i\right )}^{q+1} = -2 \Sigma O_{\left (i\right )}^{q+1}+ \Box \delta_\Sigma O_{\left (i\right )}^{q} + \left(D-2\right) K_{\left (i\right )}^{q} \,. \label{deltasigmaOq}
\end{equation}
The second identity yields
\begin{equation}
K_{\left (i\right )}^{q+1} \simeq \left(\Box+\frac{D-2}{\ell^2}\right) K_{\left (i\right )}^{q} -\frac{2}{\ell^2}\Sigma \delta_\Sigma O_{\left (i\right )}^{q+1} \label{Krecursive} \,.
\end{equation}
Finally,
\begin{equation}
\Box \left(\Sigma O_{\left (i\right )}^{q}\right) \simeq \frac{D}{\ell^2} \Sigma O_{\left (i\right )}^{q} + \Sigma O_{\left (i\right )}^{q+1} + 2 K_{\left (i\right )}^{q} \,.
\label{boxsigmaOq}
\end{equation}
Repeated application  of Eqs.~\eqref{K0}-~\eqref{boxsigmaOq} shows that the variation of the scalar density $B_{\left (i\right )}^{q}=\sqrt{\vert g \vert}O_{\left (i\right )}^{q}$ involves only descendants of order $r\leq q$. It therefore takes the triangular matrix form
\begin{equation}
\delta _{\Sigma } B_{\left (i\right )}^{q} \doteq \sum_{r=0}^{q} M_{qr}^{\left(i\right)} \ell^{-2 \left(q-r\right)} \Sigma B_{\left (i\right )}^{r} \,, \quad M_{qq}^{\left(i\right)} = D-2i-2q \,.
\label{weylBiq}
\end{equation}
The explicit derivation of these expressions is given in Appendix~\ref{recursive}. The triangular transformation law suggests the construction of operators, involving combinations of Laplacian descendants, that transform homogeneously. To do so, consider a scalar $f$ satisfying
\begin{equation}
\delta_\Sigma f\simeq w\Sigma f,\qquad
\nabla^\mu\Sigma\nabla_\mu f\simeq\frac{w}{\ell^2}\Sigma f \,.
\label{scalar-seeds}
\end{equation}
Using $\Box\Sigma\simeq D\ell^{-2}\Sigma$ and
Eqs.~\eqref{deltasigmaOq}--\eqref{boxsigmaOq} gives
\begin{align}
 \delta_\Sigma(\Box f)
 &\simeq(w-2)\Sigma\Box f
 +\frac{2w(D+w-1)}{\ell^2}\Sigma f \,, \label{deltasigmaboxf}\\
 \nabla^\mu\Sigma\nabla_\mu(\Box f)
 &\simeq\frac{w-2}{\ell^2}\Sigma\Box f
 +\frac{2w(D+w-1)}{\ell^4}\Sigma f.
 \label{one-step}
\end{align}
The presence of the non-homogeneous term in Eq.~\eqref{deltasigmaboxf} suggests the form of the differential operator that induces a homogeneous transformation when applied to a scalar $f$. Indeed, the shifted Laplacian
\begin{equation}
 Q_w=\Box-\frac{w(D+w-1)}{\ell^2}
 \label{shifted-laplacian}
\end{equation}
maps both identities in Eq.~\eqref{scalar-seeds} to the same identities
with weight $w-2$:
\begin{equation}
\delta_\Sigma(Q_wf)\simeq(w-2)\Sigma Q_wf \,,
\qquad
\nabla^\mu\Sigma\,\nabla_\mu(Q_wf)
\simeq\frac{w-2}{\ell^2}\Sigma Q_wf \,.
\end{equation}
This property allows the organization of the standard Laplacian operators in terms of $Q_w$. Starting from $\Phi_{\left(i\right)}$, we may therefore define
\begin{equation}
 \Psi_{i,0}=\Phi_{\left(i\right)},\qquad \Psi_{i,q+1}= Q_{w_q} \Psi_{i,q} \,, \qquad w_q=-2 \left(i+q\right) \,,
 \label{psi-recursion1}
\end{equation}
where the weight $w_q$ of the descendants is determined by the seed identities~\eqref{Phiidimension} and~\eqref{delsigmadelphi}. Since $Q_w$ lowers the weight by two and preserves the gradient identity, induction gives
\begin{equation}
 \delta_\Sigma\Psi_{i,q}\simeq-2(i+q)\Sigma\Psi_{i,q} \,,\qquad
 \nabla^\mu\Sigma\nabla_\mu\Psi_{i,q}
 \simeq-\frac{2(i+q)}{\ell^2}\Sigma\Psi_{i,q} \,.
 \label{psi-identities}
\end{equation}
Substituting $w_q$ into the definition of the operator $Q_w$ gives the explicit recursion
\begin{equation}
 \Psi_{i,q+1}=
 \left[\Box+\frac{2(i+q)(D-2i-2q-1)}{\ell^2}\right]\Psi_{i,q} \,.
 \label{psi-recursion}
\end{equation}
Namely, the last equation is a recursive definition chosen to preserve the two identities given in Eq.~\eqref{psi-identities}. Its density
transforms homogeneously,
\begin{equation}
\delta_\Sigma\bigl(\sqrt{|g|}\,\Psi_{i,q}\bigr)
 \doteq(D-2i-2q)\Sigma\sqrt{|g|}\,\Psi_{i,q} \,.
 \label{rw5:psi-density}
\end{equation}
In $D=2n$, the density weight vanishes at the critical order $q_{\mathrm{crit}}=n-i$. A restricted Weyl-invariant combination therefore exists within the finite descendant basis
\begin{equation}
\mathcal{A}_i = \left\{\Phi_{\left(i\right)}, \Box \Phi_{\left(i\right)}, \Box^2 \Phi_{\left(i\right)}, \ldots, \Box^{n-i} \Phi_{\left(i\right)}\right\} \,.   
\label{Aiarbitrary}
\end{equation}
For $2\leq i\leq n-1$, the critical combination admits the factorized expression for
\begin{equation}
 \Psi_{i,n-i}=\mathcal P_i(\Box)\Phi_{\left(i\right)} \,,\qquad
 \mathcal P_i(t)=\prod_{a=i}^{n-1}
 \left[t+\frac{2a(2n-2a-1)}{\ell^2}\right] \,,
 \label{critical-polynomial}
\end{equation}
where $a=i+q$, with a non-vanishing constant coefficient:
\begin{equation}
 p_{i,0}\equiv\mathcal P_i(0)
 =\frac{(n-1)!(2n-2i)!}
 {(i-1)!(n-i)!\,\ell^{2(n-i)}}>0 \,.
 \label{constant}
\end{equation}
Hence, we can normalize the critical combination as
\begin{equation}
\frac{\Psi_{i,n-i}}{p_{i,0}}
=
\Phi_{\left(i\right)}
+\sum_{q=1}^{n-i}c_{i,q}O_{\left (i\right )}^{q} \,.
\label{sumcovariantderivativecomple}
\end{equation}
Moving the curvature polynomials $\Phi_{\left(i\right)}$ to the left-hand side, isolates a sum of covariant total derivatives. Since the normalized critical density has vanishing restricted Weyl variation, this sum satisfies
\begin{equation}
\delta_\Sigma \left(\sqrt{|g|}\sum_{q=1}^{n-i}c_{i,q}O_q^{\left (i\right )}\right)= \delta_\Sigma\!\left[\sqrt{|g|}
 \left(\frac{\Psi_{i,n-i}}{p_{i,0}}-\Phi_{\left(i\right)}\right)\right]
 \doteq-2(n-i)\Sigma\sqrt{|g|}\,\Phi_{\left(i\right)} \,.
 \label{obstruction-cancellation}
\end{equation}
Therefore an obstruction of the form
\begin{equation}
C_i\,\Sigma\sqrt{|g|}\Phi_{\left(i\right)}
\end{equation}
is cancelled by the term
\begin{equation}
\frac{C_i}{2\left(n-i\right)}\sqrt{|g|}\left(\frac{\Psi_{i,n-i}}{p_{i,0}}-\Phi_{\left(i\right)}\right) \,.
\label{genericCi}
\end{equation}
This relation allows the construction of the compensating total derivative explicitly in each curvature sector and establishes the existence of the completion within the finite descendant basis.

\subsection{Closed form completion}
\label{closedcompletion}

Based on this analysis we now determine the compensating terms to the obstruction of the topologically renormalized density given in Eq.~\eqref{weylvariationIktphi}. In this case, the obstruction coefficients are $C_i=2(n-i)\beta_i$. Applying the sectorwise completion~\eqref{genericCi} therefore gives
\begin{equation}
 \mathcal L^{(2n)}
 =\mathcal L_{\mathrm{top}}^{(2n)}
 +\sqrt{|g|}\sum_{i=2}^{n-1}\beta_i
 \left(\frac{\Psi_{i,n-i}}{p_{i,0}}-\Phi_{\left(i\right)}\right) \,.
 \label{completion}
\end{equation}
The added terms are covariant total derivatives by Eq.~\eqref{sumcovariantderivativecomple}, while the sector $i=n$ already has vanishing density weight and requires no derivative completion. Defining $\Psi_{n,0}=\Phi_{\left(n\right)}$ and $p_{n,0}=1$, the density of Eq.~\eqref{completion} becomes
\begin{equation}
\mathcal L^{(2n)}
=
\sqrt{|g|}\sum_{i=2}^{n}
\frac{\beta_i}{p_{i,0}}\Psi_{i,n-i} \,.
\end{equation}
Using
\begin{equation}
\frac{\beta_i}{p_{i,0}}
=
\frac{(-1)^i\ell^{2n-2}}
{i\,2^i(2n-2)!} \,,
\end{equation}
we obtain the closed-form expression
\begin{equation}
\mathcal L^{(2n)}
 =\frac{\ell^{2n-2}\sqrt{|g|}}{(2n-2)!}
 \sum_{i=2}^{n}\frac{(-1)^i}{i\,2^i}
 \left[\prod_{a=i}^{n-1}
 \left(\Box+\frac{2a(2n-2a-1)}{\ell^2}\right)\right]\Phi_{\left(i\right)} \,.
 \label{factorized-density}
\end{equation}
The empty product for $i=n$ is understood as the identity. Eq.~\eqref{factorized-density} gives the restricted Weyl-completed density in every even bulk dimension. Expanding the factors reproduces the topological polynomial together with the required total-derivative terms
\begin{equation}
 \mathcal L^{(2n)}
 =\mathcal L_{\mathrm{top}}^{(2n)}
 +\sqrt{|g|}\sum_{k=3}^{n}\ell^{2k-2}\mathcal W_{\left(k\right)},\qquad
 \mathcal W_{\left(k\right)}=\sum_{i=2}^{k-1}a_{k-i,i}\Box^{k-i}\Phi_{\left(i\right)} \,.
 \label{W-ansatz}
\end{equation}
All coefficients $a_{k-i,i}$ are determined directly by expanding the product of Eq.~\eqref{factorized-density}. Even though they depend on the bulk dimension, this remains implicit at fixed $D=2n$.

Therefore, the candidate renormalized EAdS action provided by this prescription reads
\begin{equation}
 I^{(2n)}_{\mathrm{ren}}
 =\frac{1}{16\pi G_N}\int_{\mathcal M_{2n}}d^{2n}X\,
 \mathcal L^{(2n)}
 -\frac{c_{2n}}{16\pi G_N}(4\pi)^n n!\,\chi(\mathcal M_{2n}) \,.
 \label{action}
\end{equation}
where the additive constant is fixed by the normalization of topological renormalization adopted in Sec.~\ref{TopologicalRenormalization}.

It is important to note that the density~\eqref{factorized-density} admits a direct comparison with the ambient construction of Case et al.~\cite{Case:2024renormalized}. On Einstein--AdS backgrounds, their Proposition~3.7 yields the same shifted-Laplacian factors $Q_w$. Accounting for the normalization of the antisymmetrized curvature polynomials and combining their renormalized-volume identity with the Euler-characteristic term in the completed action yields, with the action convention adopted here,
\begin{equation}
I_{\mathrm{ren}}^{(2n)}
=-\frac{2n-1}{8\pi G_N\ell^2}
\operatorname{Vol}_{\mathrm{ren}}(M) \,.
\end{equation}
This establishes on-shell finiteness and identifies the completed action with the standard renormalized-volume expression on Poincaré--Einstein manifolds satisfying the hypotheses of that theorem, in arbitrary even dimensions.

\subsection{Application in eight and ten dimensions}
\label{examples}

\subsubsection{Eight dimensions}
\label{eight}

For $D=8$, Eq.~\eqref{factorized-density} induces quadratic and cubic curvature sectors that reach their critical orders at
$q_{\mathrm{crit}}=2$ and $q_{\mathrm{crit}}=1$, respectively. Their factorized combinations are
\begin{equation}
 \Psi_{2,2}=
 \left(\Box+\frac{12}{\ell^2}\right)
 \left(\Box+\frac{6}{\ell^2}\right)\Phi_{\left(2\right)} \,,\qquad
 \Psi_{3,1}=\left(\Box+\frac{6}{\ell^2}\right)\Phi_{\left(3\right)}\,.
 \label{eight-factors}
\end{equation}
Expanding Eq.~\eqref{factorized-density} yields
\begin{equation}
 \mathcal L^{(8)}
 =\frac{\ell^6\sqrt{|g|}}{1920}
 \left(
 \frac{24}{\ell^4}\Phi_{\left(2\right)}-\frac{2}{3\ell^2}\Phi_{\left(3\right)}
 +\frac{1}{24}\Phi_{\left(4\right)}
 +\frac{1}{3}\Box^2\Phi_{\left(2\right)}-\frac{1}{9}\Box\Phi_{\left(3\right)}
 +\frac{6}{\ell^2}\Box\Phi_{\left(2\right)}\right) \,.
 \label{L8-poly}
\end{equation}
The corresponding action reads
\begin{equation}
 I_{\mathrm{ren}}^{(8)}
 =\frac{1}{16\pi G_N}\int_{\mathcal M_8}d^8X\,
 \mathcal L^{(8)}
 -\frac{2\pi^3\ell^6}{15G_N}\chi(\mathcal M_8) \,,
\label{8DweylcounterPhi}
\end{equation}
or, retaining the Einstein--Hilbert and Euler terms explicitly,
\begin{align}
 I_{\mathrm{ren}}^{(8)}
 ={}&\frac{1}{16\pi G_N}\int_{\mathcal M_8}d^8X\sqrt{|g|}
 \left(R-2\Lambda+c_8\mathcal E_8
 +\frac{\ell^4}{320}\Box\Phi_{\left(2\right)}
 +\frac{\ell^6}{5760}\Box^2\Phi_{\left(2\right)}
 -\frac{\ell^6}{17280}\Box\Phi_{\left(3\right)}\right)\nonumber\\
 &-\frac{2\pi^3\ell^6}{15G_N}\chi(\mathcal M_8) \,.
\label{8Dweylcounter}
\end{align}
Hence, we recover the 8D conformally renormalized Einstein–AdS action~\cite{Anastasiou:2026jrt}. Its infrared finiteness therefore provides an independent check of this prescription and the resulting closed-form density~\eqref{factorized-density}.

Unlike the 6D case, the resulting action of Eq.~\eqref{8Dweylcounter} is not unique: one may add two conformally invariant total derivative terms which remain unchanged under the restricted Weyl completion. Such invariants were classified in Ref.~\cite{Boulanger:2004zf} and its corrigendum~\cite{Boulanger_2022}, and their relation to the Einstein sector was analyzed in Ref.~\cite{Boulanger:2025oli}. Interestingly, these two total derivative terms fall off faster than the normalizable order, such that their contribution vanishes as the cutoff is removed~\cite{Anastasiou:2026jrt}.

\subsubsection{Ten dimensions}
\label{ten}

For $D=10$, the quadratic, cubic, and quartic sectors terminate at
$q_{\mathrm{crit}}=3$, $q_{\mathrm{crit}}=2$, and $q_{\mathrm{crit}}=1$. The critical combinations are
\begin{align}
 \Psi_{2,3}
 &=\left(\Box+\frac{20}{\ell^2}\right)
 \left(\Box+\frac{18}{\ell^2}\right)
 \left(\Box+\frac{8}{\ell^2}\right)\Phi_{\left(2\right)} \,,\nonumber\\
 \Psi_{3,2}
 &=\left(\Box+\frac{18}{\ell^2}\right)
 \left(\Box+\frac{8}{\ell^2}\right)\Phi_{\left(3\right)} \,,\nonumber\\
 \Psi_{4,1}
 &=\left(\Box+\frac{8}{\ell^2}\right)\Phi_{\left(4\right)} \,.
 \label{ten-factors}
\end{align}
Their expansion admits the following derivative contributions
\begin{align}
 \mathcal W_{(3)}&=\frac{83}{40320}\Box\Phi_{\left(2\right)} \,,\nonumber\\
 \mathcal W_{\left(4\right)}&=\frac{23}{161280}\Box^2\Phi_{\left(2\right)}
 -\frac{13}{483840}\Box\Phi_{\left(3\right)} \,,\nonumber\\
 \mathcal W_{(5)}&=\frac{1}{322560}\Box^3\Phi_{\left(2\right)}
 -\frac{1}{967680}\Box^2\Phi_{\left(3\right)}
 +\frac{1}{2580480}\Box\Phi_{\left(4\right)} \,,
 \label{ten-W}
\end{align}
whereas the polynomial part reads
\begin{equation}
 P_{10}(F)=\frac{1}{112\ell^6}\Phi_{\left(2\right)}
 -\frac{1}{6720\ell^4}\Phi_{\left(3\right)}
 +\frac{1}{322560\ell^2}\Phi_{\left(4\right)}
 -\frac{1}{6451200}\Phi_{\left(5\right)} \,.
 \label{P10}
\end{equation}
Finally, the Weyl-completed action is
\begin{align}
 I_{\mathrm{ren}}^{(10)}
 ={}&\frac{\ell^8}{16\pi G_N}\int_{\mathcal M_{10}}d^{10}X\sqrt{|g|}
 \Biggl[P_{10}(F)
 +\frac{83}{40320\ell^4}\Box\Phi_{\left(2\right)}
 +\frac{23}{161280\ell^2}\Box^2\Phi_{\left(2\right)}
 +\frac{1}{322560}\Box^3\Phi_{\left(2\right)}\nonumber\\
 &\hspace{5em}
 -\frac{13}{483840\ell^2}\Box\Phi_{\left(3\right)}
 -\frac{1}{967680}\Box^2\Phi_{\left(3\right)}
 +\frac{1}{2580480}\Box\Phi_{\left(4\right)}\Biggr]
 +\frac{4\pi^4\ell^8}{105G_N}\chi(\mathcal M_{10}) \,,
 \label{I10-poly}
\end{align}
or equivalently
\begin{align}
 I_{\mathrm{ren}}^{(10)}
 ={}&\frac{1}{16\pi G_N}\int_{\mathcal M_{10}}d^{10}X\sqrt{|g|}
 \Biggl[R-2\Lambda-\frac{\ell^8}{201600}\mathcal E_{10}
 +\frac{83\ell^4}{40320}\Box\Phi_{\left(2\right)}
 +\frac{23\ell^6}{161280}\Box^2\Phi_{\left(2\right)}
 +\frac{\ell^8}{322560}\Box^3\Phi_{\left(2\right)}\nonumber\\
 &\hspace{5em}
 -\frac{13\ell^6}{483840}\Box\Phi_{\left(3\right)}
 -\frac{\ell^8}{967680}\Box^2\Phi_{\left(3\right)}
 +\frac{\ell^8}{2580480}\Box\Phi_{\left(4\right)}\Biggr]
 +\frac{4\pi^4\ell^8}{105G_N}\chi(\mathcal M_{10}) \,.
 \label{I10}
\end{align}
These explicit
expressions illustrate the general factorized construction. A general comparison of the ten-dimensional completion with standard holographic renormalization for arbitrary boundary data remains to be established. Nevertheless, in the next section, we evaluate the completed action on black hole solutions with non-ACF asymptotics, demonstrate the cancellation of infrared divergences, and verify agreement with standard counterterm results.

\section{Applications to topological black holes}
\label{topologicalBH}

We now test the eight- and ten-dimensional renormalized actions constructed in Sec.~\ref{evenD} on Euclidean Schwarzschild--AdS black holes with non-conformally-flat Einstein horizons. Unlike black holes with constant-curvature horizons, such as $\mathcal B_{D-2} = \mathbb{S}^{D-2}, \mathbb{H}^{D-2}, \mathbb{T}^{D-2}$, the product Einstein geometries considered below lie outside the ACF class for which topological renormalization is already sufficient~\cite{Anastasiou:2020zwc}. For these geometries, the bulk Weyl tensor squared scales as $\left \vert W\right \vert ^2 \sim r^{-4}\left \vert \widehat{W} \right \vert  ^2$, spoiling the ACF condition~\cite{Awad:2000gg,Anastasiou:2023oro}. In what follows, hatted quantities denote intrinsic tensors of the base manifold. Their evaluation allows us to examine explicitly the cancellation of IR divergences by the bulk-covariant counterterms and to compare the resulting finite actions with the HR.

We consider the metric
\begin{equation}
ds^{2}=f_D(r)d\tau^{2}
+\frac{dr^{2}}{f_D(r)}
+r^{2}d\Sigma_{\mathcal B_{D-2}}^{2} \,,
\label{eq:topological-BH-metric}
\end{equation}
where $\mathcal B_{D-2}$ is a compact Einstein manifold satisfying
\begin{equation}
\widehat R_{ij}= \lambda \widehat{g}_{ij} \,,
\label{intrinsicbaseRicci}
\end{equation}
where $\widehat{g}_{ij}$ is the intrinsic metric. The bulk Einstein equations
\begin{equation}
R_{\mu\nu}=-\frac{D-1}{\ell^{2}}g_{\mu\nu}
\end{equation}
determine
\begin{equation}
f_D(r)=
\frac{r^{2}}{\ell^{2}}
+\frac{\widehat{g}^{ij} \widehat R_{ij}}{\left(D-2\right)\left(D-3\right)}
-\frac{2m}{r^{D-3}}.
\label{eq:topological-BH-function}
\end{equation}
Thus, the curvature normalization $\lambda$ of the Einstein horizon determines the constant term in the metric function $f_D(r)$. Throughout the following examples, we adopt $\lambda=1$, following the conventions of Ref.~\cite{Clarkson:2002uj}.

These geometries belong to the class of AdS black holes with Einstein horizons discussed in Ref.~\cite{Birmingham:1998nr}. They can also be obtained by setting the NUT charge $N$ to zero in the higher-dimensional solutions of Ref.~\cite{Clarkson:2002uj}. In that case, the Misner-string condition on the Euclidean period disappears, and the period must be determined from regularity at the largest positive zero of $f_D(r)$, that is the outer horizon $r=r_{+}$, as
\begin{equation}
\beta = \frac{4 \pi}{f^{'}_D(r_{+})} \,,
\end{equation}
where $f^{'}_D(r)=\partial_r f_D(r)$.

\subsection{Application to topological black holes in eight dimensions}
\label{sec:topological-BH-D8}

For $D=8$, we examine the two six-dimensional Einstein bases
$\mathcal B_{6}=(\mathbb{S}^{2})^{3}$ and $ \mathcal B_{6}=\mathbb{S}^{2}\times\mathbb{CP}^{2}
$. Imposing the Einstein condition in product spaces requires the matching of the normalization $\lambda$ of each subspace. The metric function then becomes
\begin{equation}
f_{8}\left(r\right)=
\frac{r^{2}}{\ell^{2}}
+\frac{1}{5}
-\frac{2m}{r^{5}},
\label{eq:f8-topological}
\end{equation}
with
\begin{equation}
m=
\frac{r_{+}^{5}\left(\ell^{2}+5r_{+}^{2}\right)}
{10\ell^{2}},
\qquad
\beta=
\frac{4\pi\ell^{2}r_{+}}
{\ell^{2}+7r_{+}^{2}}\,.
\label{D8-horizon-data}
\end{equation}
To evaluate the renormalized action, we first compute the antisymmetrized AdS-curvature polynomials $\Phi_{\left(k\right)}$ defined in Eq.~\eqref{phiF}, together with their Laplacian descendants. Since $F^{\alpha \beta}_{\mu \nu}$ coincides with the Weyl tensor on Einstein--AdS configurations, its tracelessness ensures that $\Phi_{\left(1\right)}=0$. The nontrivial contributions therefore begin at quadratic order in the AdS curvature.

For the base $\mathcal B_{6}=(\mathbb{S}^{2})^{3}$, the transverse sector of the bulk metric~\eqref{eq:topological-BH-metric} reads
\begin{equation}
d\Sigma_{(\mathbb{S}^2)^3}^2
=
\sum_{a=1}^{3}
\left(
d\theta_a^{2}
+\sin^{2}\theta_a\,d\phi_a^{2}
\right)\,,
\label{s2cube}
\end{equation}
and the direct tensor contractions give
\begin{align}
\Phi_{\left(2\right)} \left[{(\mathbb{S}^{2})^{3}}\right]
&= \frac{192 \left(r^{10} +525 m^2\right)}{5 r^{14}}
\,, \\
\Phi_{\left(3\right)} \left[{(\mathbb{S}^{2})^{3}}\right]
&=\frac{3072 \left(2 r^{15}+45 m r^{10}+34125m^3\right)}{25 r^{21}} \,,
\\
\Phi_{\left(4\right)} \left[{(\mathbb{S}^{2})^{3}}\right]
&= \frac{73728m \left(8r^{15}+55 mr^{10}+5250 m^3\right)}{5r^{28}} \,.
\end{align}
The terms independent of $m$ in $\Phi_{\left(2\right)}$ and $\Phi_{\left(3\right)}$ originate from the intrinsic Weyl curvature of the product base. Consequently, even the $m=0$ member of this family spoils the ACF condition.

For $\mathcal B_{6}=\mathbb{S}^{2}\times\mathbb{CP}^{2}$, the base metric takes the form
\begin{align}
d\Sigma_{\mathbb{S}^2\times\mathbb{CP}^2}^{2}
&=
d\Omega_2^{2}
+d\Sigma_{\mathbb{CP}^2}^{2},
\qquad
d\Omega_2^2
=
d\theta_1^2+\sin^2\theta_1\,d\phi_1^2  \,, \nonumber \\ d\Sigma_{\mathbb{CP}^2}^{2}
&=
\frac{du^{2}}
{\left(1+\frac{u^{2}}{6}\right)^{2}}
+
\frac{u^{2}}
{4\left(1+\frac{u^{2}}{6}\right)^{2}}
\left(d\psi+\cos\theta\,d\phi\right)^{2}
+
\frac{u^{2}}
{4\left(1+\frac{u^{2}}{6}\right)}
\left(
d\theta^{2}
+\sin^{2}\theta\,d\phi^{2}
\right) \,.
\label{S2CP2}
\end{align}
Hence, the curvature tensor contractions give
\begin{align}
\Phi_{\left(2\right)} \left[{\mathbb{S}^{2}\times\mathbb{CP}^{2}}\right]
&=
\frac{32 \left(13r^{10} +9450 m^2\right)}{15 r^{14}}
\,,
\label{eq:Phi2-D8-S2CP2}
\\
\Phi_{\left(3\right)} \left[{\mathbb{S}^{2}\times\mathbb{CP}^{2}}\right]
&=
\frac{512 \left(7 r^{15}+195 m r^{10}+204750 m^3\right)}{25 r^{21}} \,,
\label{eq:Phi3-D8-S2CP2}
\\
\Phi_{\left(4\right)} \left[{\mathbb{S}^{2}\times\mathbb{CP}^{2}}\right]
&=
\frac{4096 m \left(84 r^{15}+715 mr^{10}+94500 m^3\right)}{5r^{28}} \,.
\label{eq:Phi4-D8-S2CP2}
\end{align}
Notice that the term containing the highest powers of the mass parameter $m$ are identical for both horizon geometries
\begin{equation}
\Phi_{\left(2\right)} =\frac{20160m^{2}}{r^{14}} \,,
\qquad
\Phi_{\left(3\right)} =\frac{4193280m^{3}}{r^{21}} \,,
\qquad
\Phi_{\left(4\right)} = \frac{77414400m^{4}}{r^{28}} \,,
\end{equation}
as they arise entirely from the mass-dependent contribution to the AdS curvature tensor, which is independent of the intrinsic geometry of the Einstein horizon. By contrast, the remaining terms involve the intrinsic Weyl curvature of the base manifold and therefore distinguish $(\mathbb{S}^2)^3$ from $\mathbb{S}^2\times\mathbb{CP}^2$. This separation between universal mass-dependent contributions and base-dependent curvature terms will also appear in the ten-dimensional examples.

Since the curvature scalars depend only on $r$, we introduce the radial Laplacian for a scalar $\Omega\left(r\right)$
\begin{equation}
\Box \Omega\left(r\right)=
\frac{1}{r^{2\left(n-1\right)}}
\frac{d}{dr}
\left[
r^{2\left(n-1\right)}f_{2n}\left(r\right)\frac{d\Omega}{dr}
\right] \,.
\label{radial-Laplacian-D2n}
\end{equation}
The expressions for
$\Box\Phi_{\left(2\right)}$, $\Box^{2}\Phi_{\left(2\right)}$, and
$\Box\Phi_{\left(3\right)}$ entering the eight-dimensional action are obtained by repeatedly applying Eq.~\eqref{radial-Laplacian-D2n} to the explicit formulas above for $n=4$.

At radial cutoff $r_c$, the regulated topologically renormalized action~\eqref{L2ntop} is
\begin{equation}
I^{(8)}_{\mathrm{top}}\left(r_c\right)=
\frac{\beta V_{\mathcal B_{6}}}
{16\pi G_N}
\int_{r_{+}}^{r_c}dr \,
r^{6}\tilde{\mathcal{L}}^{\left(8\right)}_{\textrm{top}}\left(r\right) \,,
\label{radial-action-D8}
\end{equation}
where $\tilde{\mathcal{L}}^{\left(2n\right)}_{\textrm{top}}=\frac{\mathcal{L}^{\left(2n\right)}_{\textrm{top}}}{\sqrt{g}}$. The Euler characteristic contribution then reads
\begin{equation}
I_{0}^{\left(8\right)}=
-\frac{2\pi^{3}\ell^{6}}{15G_N}\chi(\mathcal{M}_{8}) \,,
\end{equation}
where \(\chi(\mathcal{M}_{8})=\chi(\mathcal B_{6})\) for the Euclidean topology
\(\mathcal{M}_{8}\simeq \mathbb{D}_{2}\times\mathcal B_{6}\), where $\mathbb{D}_{2}$ is the thermal disc. The relevant geometric data of both cases are
\begin{equation}
\begin{aligned}
V_{(\mathbb{S}^{2})^{3}}=64\pi^{3} \,,
\quad 
\chi\big((\mathbb{S}^{2})^{3}\big)=8 \,, \quad \\
V_{\mathbb{S}^{2}\times\mathbb{CP}^{2}}=72\pi^{3}\,, \quad
\chi\big(\mathbb{S}^{2}\times\mathbb{CP}^{2}\big)=6 \,.
\end{aligned}
\end{equation}
The explicit expressions make the cancellation of divergences particularly transparent. For example,
$r^{6}\Phi_{\left(2\right)}\sim r^{2} \,,
 \quad
 r^{6}\Phi_{\left(3\right)}\sim r^{0} \,,
$
at large $r$. Their separate radial integrals are therefore divergent. Nevertheless, after substituting all the $\Phi_{\left(k\right)}$ and their Laplacian descendants into the Weyl completed action~\eqref{8DweylcounterPhi}, every positive power of $r_c$ cancels identically.

In particular, after imposing Eq.~\eqref{D8-horizon-data} and taking $r_c \rightarrow\infty$, one obtains
\begin{equation}
I^{\left(8\right)}_{\mathrm{ren}}\left[\mathcal B_{6}\right]
=
\frac{V_{\mathcal B_{6}}r_{+}^{6}
\left(5r_{+}^{2}-\ell^{2}\right)}
{20G_N\left(\ell^{2}+7r_{+}^{2}\right)} \,,
\label{Iren-D8-general}
\end{equation}
or equivalently,
\begin{align}
I^{\left(8\right)}_{\mathrm{ren}}\left[(\mathbb{S}^{2})^{3}\right]
&=
\frac{16\pi^{3}r_{+}^{6}
\left(5r_{+}^{2}-\ell^{2}\right)}
{5G_N\left(\ell^{2}+7r_{+}^{2}\right)} \,,
\label{Iren-D8-S2S2S2}
\\
I^{\left(8\right)}_{\mathrm{ren}}
\left[\mathbb{S}^{2}\times\mathbb{CP}^{2}\right]
&=
\frac{18\pi^{3}r_{+}^{6}
\left(5r_{+}^{2}-\ell^{2}\right)}
{5G_N\left(\ell^{2}+7r_{+}^{2}\right)} \,.
\label{Iren-D8-S2CP2}
\end{align}
Therefore, the action~\eqref{8DweylcounterPhi} is free of IR divergences even in the case of non-ACF boundaries.

\subsection{Application to topological black holes in ten dimensions}
\label{sec:topological-BH-D10}

For $D=10$, we consider the three Einstein bases
\begin{equation}
\mathcal B_{8}=(\mathbb{S}^{2})^{4} \,,
\qquad
\mathcal B_{8}=(\mathbb{S}^{2})^{2}\times\mathbb{CP}^{2} \,,
\qquad
\mathcal B_{8}=\mathbb{CP}^{2}\times\mathbb{CP}^{2} \,,
\end{equation}
presented in Ref.~\cite{Clarkson:2002uj}. The corresponding metrics are
\begin{align}
d\Sigma_{(\mathbb{S}^2)^4}^{2}
&=
\sum_{a=1}^{4}
\left(
d\theta_a^{2}
+\sin^{2}\theta_a\,d\phi_a^{2}
\right)\,, \\
d\Sigma_{\left(\mathbb{S}^2\right)^2 \times\mathbb{CP}^2}^{2}&= \sum_{a=1}^{2}
\left(
d\theta_a^{2}
+\sin^{2}\theta_a\,d\phi_a^{2}
\right)+ d\Sigma_{\mathbb{CP}^2}^{2} \,, \\
d\Sigma_{\mathbb{CP}^2 \times \mathbb{CP}^2}^{2} &=d\Sigma_{\mathbb{CP}^2}^{2} + d\Sigma_{\mathbb{CP}^{2'}}^{2} \,,
\end{align}
where the line element of $d\Sigma_{\mathbb{CP}^2}^{2}$ is given in Eq.~\eqref{S2CP2}.

With $\lambda=1$, the metric function reads
\begin{equation}
f_{10}\left(r\right)
=
\frac{r^{2}}{\ell^{2}}
+\frac{1}{7}
-\frac{2m}{r^{7}} \,,
\label{f10-topological}
\end{equation}
and regularity at $r=r_{+}$ gives
\begin{equation}
m
=
\frac{r_{+}^{7}
\left(\ell^{2}+7r_{+}^{2}\right)}
{14\ell^{2}} \,,
\qquad
\beta
=
\frac{4\pi\ell^{2}r_{+}}
{\ell^{2}+9r_{+}^{2}} \,.
\label{D10-horizon-data}
\end{equation}
The ten-dimensional curvature polynomials $\Phi_{\left(k\right)}$, with $2\leq k\leq5$, are considerably more involved than their eight-dimensional counterparts, and their explicit expressions are collected in Appendix~\ref{10Dcurvature}. Nevertheless, they exhibit the same characteristic separation between universal mass-dependent terms and contributions determined by the intrinsic geometry of the Einstein horizon. In particular, the terms of the highest powers of the mass parameter are independent of the chosen base manifold and take the form
\begin{equation}
\begin{aligned}
\Phi_{\left(2\right)}
&=\frac{64512m^{2}}{r^{18}} \,,
\Phi_{\left(3\right)}
=\frac{32901120m^{3}}{r^{27}} \,, \\
\Phi_{\left(4\right)}
&=\frac{3622993920m^{4}}{r^{36}} \,,
\Phi_{\left(5\right)}
=\frac{52022476800m^{5}}{r^{45}} \,.
\label{massdependent10D}
\end{aligned}
\end{equation}
The remaining terms depend on the intrinsic curvature of the base. Therefore, the three base geometries are locally distinguishable even though their final renormalized actions have the same functional dependence on $r_{+}$.

The derivative terms
$
\Box\Phi_{\left(2\right)} \,,\quad
\Box^{2}\Phi_{\left(2\right)} \,,\quad
\Box^{3}\Phi_{\left(2\right)} \,,\quad
\Box\Phi_{\left(3\right)} \,,\quad
\Box^{2}\Phi_{\left(3\right)} \,,\quad
\Box\Phi_{\left(4\right)} \,
$
are obtained by repeated application of the radial Laplacian~\eqref{radial-Laplacian-D2n} for $n=5$.
Introducing a radial cutoff $r_c$, the regulated topologically renormalized contribution in ten dimension is
\begin{equation}
I^{\left(10\right)}_{\mathrm{top}}\left(r_c\right)=
\frac{\beta V_{\mathcal B_{8}}}
{16\pi G_N}
\int_{r_{+}}^{r_c}dr \,
r^{8}\tilde{\mathcal{L}}^{(10)}_{\textrm{top}}(r) \,,
\label{radial-action-D10}
\end{equation}
whereas the corresponding topological contribution reads
\begin{equation}
I_{0}^{\left(10\right)}
=
\frac{4\pi^{4}\ell^{8}}{105G_N}\chi(M_{10}) \,,
\qquad
\chi(\mathbb{M}_{10})=\chi(\mathcal B_{8}) \,.
\end{equation}
The volumes and Euler characteristics of the eight-dimensional
Einstein bases are summarized in Table~\ref{tab:10D-base-data}.

\begin{table}[!htbp]
    \centering
    \caption{Volumes and Euler characteristics of the eight-dimensional
    Einstein base manifolds considered in the ten-dimensional solutions.
    The metrics are normalized according to
    $\widehat{R}_{ij}=\widehat g_{ij}$.}
    \label{tab:10D-base-data}
    \begin{tabular}{lcc}
        \toprule
        Base manifold $\mathcal{B}_{8}$
        & Volume $V_{\mathcal{B}_{8}}$
        & Euler characteristic $\chi(\mathcal{B}_{8})$ \\
        \midrule
        $(\mathbb{S}^{2})^{4}$
        & $256\pi^{4}$
        & $16$ \\

        $(\mathbb{S}^{2})^{2}\times\mathbb{CP}^{2}$
        & $288\pi^{4}$
        & $12$ \\

        $\mathbb{CP}^{2}\times\mathbb{CP}^{2}$
        & $324\pi^{4}$
        & $9$ \\
        \bottomrule
    \end{tabular}
\end{table}

The asymptotic behavior of the individual curvature polynomials reveals the divergent contributions to the action~\eqref{radial-action-D10}. In particular, for the different product horizon geometries, the leading-order falloffs are
\begin{equation}
r^{8}\Phi_{\left(2\right)}\sim r^{4},
\qquad
r^{8}\Phi_{\left(3\right)}\sim r^{2},
\qquad
r^{8}\Phi_{\left(4\right)}\sim r^{0} \,.
\end{equation}
Thus, their individual radial integrals contain power-law divergences. Nevertheless, when the polynomial $P_{10}(F)$ is combined with the Laplacian descendants determined by the restricted Weyl completion, all cutoff-divergent contributions cancel.

After imposing Eq.~\eqref{D10-horizon-data} and taking taking $r_c\to\infty$, the result can be written as
\begin{equation}
I^{\left(10\right)}_{\mathrm{ren}}[\mathcal B_{8}]
=
\frac{V_{\mathcal B_{8}}r_{+}^{8}
\left(7r_{+}^{2}-\ell^{2}\right)}
{28G_N\left(\ell^{2}+9r_{+}^{2}\right)} \,.
\label{Iren-D10-general}
\end{equation}
For the three bases, this gives
\begin{align}
I^{\left(10\right)}_{\mathrm{ren}}\big[(\mathbb{S}^{2})^{4}\big]
&=
\frac{64\pi^{4}r_{+}^{8}
\left(7r_{+}^{2}-\ell^{2}\right)}
{7G_N\left(\ell^{2}+9r_{+}^{2}\right)} \,,
\label{Iren-D10-S2}
\\
I^{\left(10\right)}_{\mathrm{ren}}
\left[(\mathbb{S}^{2})^{2}\times\mathbb{CP}^{2}\right]
&=
\frac{72\pi^{4}r_{+}^{8}
\left(7r_{+}^{2}-\ell^{2}\right)}
{7G_N\left(\ell^{2}+9r_{+}^{2}\right)} \,,
\label{Iren-D10-S2S2CP2}
\\
I^{\left(10\right)}_{\mathrm{ren}}
\left[\mathbb{CP}^{2}\times\mathbb{CP}^{2}\right]
&=
\frac{81\pi^{4}r_{+}^{8}
\left(7r_{+}^{2}-\ell^{2}\right)}
{7G_N\left(\ell^{2}+9r_{+}^{2}\right)} \,.
\label{Iren-D10-CP2CP2}
\end{align}
The resulting renormalized action therefore provides an explicit test of the ten-dimensional construction for Einstein-horizon geometries that induce non-ACF spacetimes.

\subsection{Comparison with the standard counterterm result}

To assess the consistency of the restricted Weyl prescription, we compare the finite actions obtained above with the standard holographic counterterm result. For this purpose, we consider the vanishing NUT-charge limit $N=0$ of the higher-dimensional solutions analyzed by  Clarkson, Fatibene and Mann~\cite{Clarkson:2002uj} using the standard counterterm prescription~\cite{Emparan:1999pm,deHaro:2000vlm}. Restoring $G_N$ and replacing the product-sphere volume by the corresponding volume $V_{\mathcal B_{D-2}}$, the resulting Euclidean action takes the form
\begin{equation}
\left.I_{E}^{\mathrm{CFM}}\right|_{N=0}
=
\frac{\beta V_{\mathcal B_{D-2}}}
{8\pi G_N\ell^{2}}
\left(m\ell^{2}-r_{+}^{D-1}\right) \,.
\label{CFM-N0}
\end{equation}
Using the horizon condition and the regular Euclidean period, we obtain
\begin{equation}
\left.I_{E}^{\mathrm{CFM}}\right|_{N=0}
=
\frac{V_{\mathcal B_{D-2}}r_{+}^{D-2}
\left[\ell^{2}-(D-3)r_{+}^{2}\right]}
{4(D-3)G_N
\left[\ell^{2}+(D-1)r_{+}^{2}\right]} \,.
\label{CFM-N0-horizon}
\end{equation}
Because the Euclidean action convention adopted in Ref.~\cite{Clarkson:2002uj} differs by an overall sign  from the convention used here, the two results are related by
\begin{equation}
I^{(D)}_{\mathrm{ren}}
=
-\left.I_{E}^{\mathrm{CFM}}\right|_{N=0} \,.
\end{equation}
The agreement with the standard counterterm result provides a nontrivial consistency check of the restricted Weyl prescription. The polynomials $\Phi_{\left(i\right)}$ depend on the local geometry of the Einstein horizon, and their individual integrals contain infrared divergences, except for the top-degree invariant $\Phi_{\left(n\right)}$ in $D=2n$. These divergences cancel in the completed action once the required Laplacian descendants are included. The resulting finite on-shell actions depend on the horizon geometry only through its volume, in agreement with the thermodynamic expression obtained by standard holographic renormalization. This confirms the effectiveness of the symmetry-based counterterm completion for the non-ACF black-hole configurations considered here.

\section{Discussion}
\label{discussion}

We have developed a symmetry-based prescription for the construction of bulk-covariant counterterms in even-dimensional Einstein–AdS gravity. The scheme uses infinitesimal Weyl transformations that preserve the  Einstein–AdS equations asymptotically~\eqref{asymptEAdS} and requires the restricted Weyl variation of the completed density to vanish modulo terms whose integrals remain finite as the radial cutoff is removed~\eqref{variationdiv}. This criterion fixes the Euler density coupling and determines compensating covariant total derivatives using the symmetry condition~\eqref{weyleinsteincondition}, without imposing infrared finiteness directly on the regulated action.

Variations along the scaling function $\Sigma (X)$ induce recursive identities~\eqref{deltasigmaOq}--\eqref{boxsigmaOq} applicable to each curvature sector $\Phi_{\left(i\right)}$. Most importantly, these recursive relations admit combinations with definite Weyl weights~\eqref{psi-identities} which are constructed as products of shifted Laplacians $Q_w$. Their density weight vanishes at $q_{\mathrm{crit}}=n-i$, providing an explicit completion within a finite descendant basis in arbitrary even dimensions.

The resulting expression reproduces the known conformally renormalized Einstein–AdS actions in four~\cite{Anastasiou:2016jix}, six~\cite{Anastasiou:2020mik} and eight dimensions~\cite{Anastasiou:2026jrt} and yields the ten-dimensional completion involving derivatives up to $\Box^3\Phi_{\left(2\right)}$.

The eight- and ten-dimensional black-hole applications in Sec.~\ref{topologicalBH} provide independent tests of this prescription. Their product-space horizons induce non-conformally-flat boundary geometries for which topological renormalization alone is insufficient~\cite{Anastasiou:2020zwc,Anastasiou:2023oro}. The total derivative terms determined by restricted Weyl completion cancel the remaining divergences and yield finite on-shell actions in agreement with standard holographic renormalization.

The comparison with the ambient construction of Ref.~\cite{Case:2024renormalized} gives a general geometric interpretation of these results. On Einstein backgrounds, the factorized combinations agree, after matching normalizations, with the corresponding ambient invariants. Their renormalized-volume identity establishes on-shell finiteness on Poincaré--Einstein manifolds in arbitrary even dimensions. Although the leading Weyl parameter used in the derivation must satisfy the almost--Einstein scale equation, the completed action can be evaluated on any AlAdS-Einstein background covered by that theorem.

An important next step is to extend the variational analysis to arbitrary even dimensions, establishing a Dirichlet variational principle for the completed action with respect to the conformal boundary metric $g_{\left(0\right)ij}$ and determining the associated holographic stress tensor. Future work could extend the construction to higher-curvature theories and incorporate matter couplings. Another important direction is the extension to odd bulk dimensions, where logarithmic terms in the FG expansion and the boundary Weyl anomaly suggest replacing strict asymptotic invariance by an anomaly-compatible notion of Weyl covariance. More generally, the connection between restricted Weyl transformations and the conformal geometry of Einstein spaces may provide a useful framework for organizing renormalization in other gravitational settings.

\section*{Acknowledgements}

G.A.~thanks I.J.~Araya, C.~Corral,  D.~Diaz, A. Mukhopadhyay and R.~Olea for interesting discussions. G.A.~thanks the Instituto de F\'isica of the Pontificia Universidad Cat\'olica de Valpara\'iso and the Instituto de Ciencias Exactas y Naturales de la Universidad Arturo Prat for their hospitality during the early stages of this work. This work was partially funded by Agencia Nacional de Investigación y Desarrollo (ANID) through the FONDECYT grants No.~11240059, 1240043, and 1261016.

\appendix

\section{Radial foliation and Fefferman--Graham conventions} \label{GC}

We collect here the radial-foliation conventions and Fefferman--Graham expansions used in Section~\ref{FGweyl}. For the radial decomposition of geometric tensors, we write the metric in Gaussian coordinates as
\begin{equation}
ds^2 = N^2 \left(z\right) d{z^2} + h_{ij} \left(z, x^i\right) d{x^i} d{x^j} \,,
\end{equation}
where $N\left(z\right)$ is the lapse function. The normal to the constant-$z$ hypersurfaces takes the form
\begin{equation}
    n_{\mu} = -N \delta^z_{\mu} \,.
\end{equation}
The extrinsic curvature is defined by
\begin{equation}
K_{ij} = - \frac{1}{2N} \partial_z h_{ij} \,.
\end{equation}
The independent components of the Christoffel symbols are written as
\begin{equation}
\Gamma^z_{zz} =N^{-1}\partial_zN \,, \quad \Gamma^z_{ij} = \frac{1}{N} K_{ij} \,, \quad \Gamma^{i}_{zj} = -N K_j^i \,, \quad \Gamma^i_{jk}\left(g\right) = \Gamma^i_{jk} \left(h\right) \,.
\end{equation}
The above expressions lead to the Gauss-Codazzi relations for a radial foliation of the spacetime
\begin{align}
R^{ij}_{km} &=\mathcal{R}^{ij}_{km} \left(h\right)- \left(K^i_k K^j_m -K^i_m K^j_k\right) \,, \\
R^{iz}_{mz}&=\frac{1}{N} \partial_z K^i_m - K^i_s K^s_m \,, \\
R^{iz}_{jm}&=-\frac{1}{N} \left(\mathcal{D}_j K^i_m - \mathcal{D}_m K^i_j\right) \,, \\
R&=\mathcal{R}\left(h\right) - \left(K^2 + K^i_j K^j_i\right) +\frac{2}{N} \partial_z K \,,
\end{align}
where $\mathcal{D}_j$ is the covariant derivative compatible with the metric $h_{ij}$.
Specializing to FG gauge, $N=\frac{\ell}{z}$ and $h_{ij}=\frac{\ell^2}{z^2} \gamma_{ij}$, the relevant Christoffel symbols for EAdS spacetimes become
\begin{align}
\Gamma^m_{ij}&=\Gamma^m_{\left(0\right)ij} + \frac{z^2}{2 \ell^2}\left(\mathcal{D}_{\left(0\right)j} g^m_{\left(2\right)i}+\mathcal{D}_{\left(0\right)i} g^m_{\left(2\right)j}-\mathcal{D}^m_{\left(0\right)} g_{\left(2\right)ij}\right)+ \mathcal{O}\left(z^4\right) \,, \\
\Gamma^i_{zj}&=-\frac{1}{z} \delta^i_j +\frac{z}{\ell^2} g^i_{\left(2\right)j}+\mathcal{O}\left(z^3\right) \,, \\
\Gamma^z_{ij}&=\frac{1}{z} g_{\left(0\right)ij}-\frac{z^3}{\ell^4} g_{\left(4\right)ij}+ \mathcal{O}\left(z^5\right) \,, \\
\Gamma^z_{zz}&=-\frac{1}{z} \,.
\end{align}
Taking into account the FG expansion in Eq.~\eqref{FGexpansion}, we determine the asymptotic behavior of the Weyl tensor components as
\begin{align}
W^{iz}_{jz} &= \frac{z^2}{\left(d-1\right)\ell^2} \left\{-\left(\mathcal{R}^{\left(tf\right)}_{\left(0\right)}\right)^{i}_{j}+\frac{1}{\ell^2} \left[-\left(d-2\right) \left(g^{\left(tf\right)}_{\left(2\right)}\right)^i_j +\frac{1}{4} g_{\left(1\right)}\left(g^{\left(tf\right)}_{\left(1\right)}\right)^i_j +\frac{d-3}{4} \left(g^{2}_{\left(1\right)}\right)^{\left(tf\right)i}_{j}\right]\right\}+\mathcal{O}\left(z^3\right) \,, \nonumber \\
W^{iz}_{ms} &=\frac{z^2}{\ell^3} \left[\mathcal{D}_{\left(0\right)[m} g^{i}_{\left(1\right)s]}+ \frac{1}{d-1}\left(\delta^{i}_{[m} \mathcal{D}_{\left(0\right)s]}g_{\left(1\right)} -\delta^{i}_{[m} \mathcal{D}_{\left(0\right)a}g^{a}_{\left(1\right)s]} \right)\right] +\mathcal{O}\left(z^3\right) \,, \nonumber \\
W^{ij}_{ms} &=-\frac{z^2}{\left(d-1\right)\ell^4} \left[\frac{3}{4d}\operatorname{tr}\left(g_{\left(1\right)}^2\right) \delta^{ij}_{ms} -\frac{1}{4d} g^{2}_{\left(1\right)} \delta^{ij}_{ms} +g_{\left(1\right)} g^{[i}_{\left(1\right) [m} \delta^{j]}_{s]} -2\delta^{[i}_{[m} g^{j]}_{\left(1\right)a} g^{a}_{\left(1\right)s]} - \frac{d-1}{2}g^{[i}_{\left(1\right)[m}g^{j]}_{\left(1\right)s]}\right. \nonumber\\
&\left.-\frac{2}{d}g_{\left(2\right)} \delta^{ij}_{ms} +4\delta^{[i}_{[m}g^{j]}_{\left(2\right)s]} - 4 \ell^2 \delta^{[i}_{[m} \mathcal{R}^{j]}_{\left(0\right)s]}+\frac{1}{d} \ell^2 \mathcal{R}_{\left(0\right)} \delta^{ij}_{ms} +\left(d-1\right)\ell^{2} \mathcal{R}^{ij}_{\left(0\right) ms}\right]  +\mathcal{O}\left(z^3\right)  \,,
\label{weylcomponents}
\end{align}
where $\left(tf\right)$ denotes the trace-free part of the corresponding tensor. Indices are raised with the inverse conformal boundary metric $g_{\left(0\right)}^{ij}$. We denote the traces of the FG coefficients by $g_{\left(k\right)}=g_{\left(0\right)}^{ij}g_{\left(k\right)ij}$ and write
\begin{equation}
\operatorname{tr}\left(g_{\left(1\right)}^2\right)
= g_{\left(1\right)ij}g_{\left(1\right)}^{ij} \,.
\end{equation}

\section{Infinitesimal Weyl variation of the Euler density}
\label{Eulerweylvariation}

For completeness, we derive here the infinitesimal conformal variation of the Euler density for an unrestricted scaling function $\sigma \left(X\right)$. The restriction $\sigma\rightarrow \Sigma$ is imposed only in Sec.~\ref{TopologicalRenormalization}. 

In $D=2n$ dimensions, its variation reads
\begin{align}
\delta _{\sigma } \left(\sqrt{\lvert g\rvert} \mathcal{E}_{2n}\right)&=\dfrac{1}{2^{n}} \delta _{\sigma } \left(\sqrt{\lvert g\rvert} \delta _{\nu _{1} \ldots  \nu _{2 n}}^{\mu _{1} \ldots  \mu _{2 n}} R_{\mu _{1} \mu _{2}}^{\nu _{1} \nu _{2}} \cdots  R_{\mu _{2 n -1} \mu _{2 n}}^{\nu _{2 n -1} \nu _{2 n}}\right) \nonumber \\
&=\dfrac{n\sqrt{\lvert g\rvert}}{2^{n}} \delta _{\nu _{1} \ldots  \nu _{2 n}}^{\mu _{1} \ldots  \mu _{2 n}} R_{\mu _{1} \mu _{2}}^{\nu _{1} \nu _{2}} \cdots  R_{\mu _{2 n -3} \mu _{2 n -2}}^{\nu _{2 n -3} \nu _{2 n -2}}\left(2 \sigma  R_{\mu _{2 n -1} \mu _{2 n}}^{\nu _{2 n -1} \nu _{2 n}} 
+  \delta _{\sigma } R_{\mu _{2 n -1} \mu _{2 n}}^{\nu _{2 n -1} \nu _{2 n}} \right) \,.
\label{weylvareulerv1}
\end{align}
The variation of the Riemann tensor appearing in the last term can be written as
\begin{equation}
\delta _{\sigma } R_{\mu _{2 n -1} \mu _{2 n}}^{\nu _{2 n -1} \nu _{2 n}} = \delta _{\sigma } \left (g^{\lambda  \nu _{2 n}} R_{\;\lambda  \mu _{2 n -1} \mu _{2 n}}^{\nu _{2 n -1}}\right ) =-2 \sigma R_{\mu _{2 n -1} \mu _{2 n}}^{\nu _{2 n -1} \nu _{2 n}} + g^{\lambda  \nu _{2 n}} \delta _{\sigma } R_{\;\lambda  \mu _{2 n -1} \mu _{2 n}}^{\nu _{2 n-1}} \,.
\end{equation}
This contribution cancels part of Eq.~\eqref{weylvareulerv1}, such that
\begin{equation}
\delta _{\sigma } \left(\sqrt{\lvert g\rvert} \mathcal{E}_{2n}\right)= -\dfrac{n \sqrt{\lvert g\rvert}}{2^{n}} \delta _{\nu _{1} \ldots  \nu _{2 n}}^{\mu _{1} \ldots  \mu _{2 n}} R_{\mu _{1} \mu _{2}}^{\nu _{1} \nu _{2}} \cdots  R_{\mu _{2 n -3} \mu _{2 n -2}}^{\nu _{2 n -3} \nu _{2 n -2}} g^{\lambda  \nu _{2 n -1}} \delta _{\sigma } R_{\;\lambda  \mu _{2 n -1} \mu _{2 n}}^{\nu _{2 n}} \,.
\label{weylvareulerdenspartial}
\end{equation}
For an arbitrary variation of the Riemann tensor, the following relation holds
\begin{equation}
\delta R^\alpha_{ \ \beta \mu \nu}= \nabla_{\mu} \delta \Gamma^{\alpha}_{\beta \nu}-\nabla_{\nu} \delta \Gamma^{\alpha}_{\beta \mu} \,,
\end{equation}
where
\begin{equation}
\delta \Gamma^{\alpha}_{\mu \nu}= \frac{1}{2} g^{\alpha \sigma}\left(\nabla _{\mu}\delta g_{\sigma \nu}+ \nabla _{\nu}\delta g_{\mu  \sigma } - \nabla_{\sigma}\delta g_{\mu  \nu}\right) \,.
\end{equation}
Substituting these two relations into Eq.~\eqref{weylvareulerdenspartial} yields
\begin{align}
\delta _{\sigma } \left(\sqrt{\lvert g\rvert} \mathcal{E}_{2n}\right) = -\dfrac{n \sqrt{\lvert g\rvert}}{2^{n -2}} \delta _{\nu _{1} \ldots  \nu _{2 n -1}}^{\mu _{1} \ldots  \mu _{2 n -1}} R_{\mu _{1} \mu _{2}}^{\nu _{1} \nu _{2}} \cdots  R_{\mu _{2 n -3} \mu _{2 n -2}}^{\nu _{2 n -3} \nu _{2 n -2}}  \nabla _{\mu _{2 n -1}} \nabla ^{\nu _{2 n -1}}\sigma  \nonumber  \\
= -\dfrac{n \sqrt{\lvert g\rvert}}{2^{n -2}} \delta _{\nu _{1} \ldots  \nu _{2 n -1}}^{\mu _{1} \ldots  \mu _{2 n -1}}  \nabla _{\mu _{2 n -1}}\left (R_{\mu _{1} \mu _{2}}^{\nu _{1} \nu _{2}} \cdots  R_{\mu _{2 n -3} \mu _{2 n -2}}^{\nu _{2 n -3} \nu _{2 n -2}}  \nabla ^{\nu _{2 n -1}}\sigma \right )\,.
\end{align}

\section{Alternative derivation of the restricted Weyl condition}
\label{Weyleinsteincondition2}

In Sec.~\ref{restrictedweyl}, the restricted Weyl condition was derived by requiring preservation of the asymptotic Einstein–AdS equations. We provide an equivalent derivation by decomposing the AdS curvature into Weyl and Schouten tensors.

Under an arbitrary scaling function $\sigma=\sigma \left(X\right)$, the metric and Weyl tensors transform as
\begin{equation}
\delta _{\sigma } g_{\mu  \nu } =2 \sigma g_{\mu  \nu } \,, \quad \delta _{\sigma }W_{\mu \nu }^{\alpha \beta } = -2\sigma W_{\mu \nu }^{\alpha \beta } \,,
\label{infinitesimalWeyltensorvar}
\end{equation}
where
\begin{equation}
W_{\mu \nu }^{\alpha \beta } =R_{\mu \nu }^{\alpha \beta } -4S_{[\mu }^{[\alpha }\delta _{\nu ]}^{\beta ]} \,.
\label{weyltensor}
\end{equation}
Since we are interested in AlAdS spacetimes, it is convenient to introduce the following decomposition~\cite{Anastasiou:2020mik} :
\begin{equation}
W_{\mu \nu }^{\alpha \beta } =F_{\mu \nu }^{\alpha \beta } -Y_{\mu \nu }^{\alpha \beta } \,,
\label{Weyldecomposition}
\end{equation}
where 
\begin{equation}
Y_{\mu \nu }^{\alpha \beta }=4\left (S_{[\mu }^{[\alpha } +\frac{1}{2\ell ^{2}}\delta _{[\mu }^{[\alpha }\right )\delta _{\nu ]}^{\beta ]} \,.
\label{Ytensor}
\end{equation}
On an EAdS manifold, the Schouten tensor
\begin{equation}
S_{\mu \nu} = \frac{1}{D-2} \left[R_{\mu \nu}-\frac{1}{2 \left(D-1\right)} R g_{\mu \nu} \right] \,,
\end{equation}
\label{Schouten}
becomes proportional to the metric as
\begin{equation}
S_{\mu\nu}|_E = -\frac{1}{2\,\ell^2}\,g^{(E)}_{\mu\nu}\,.
\label{SchoutenEinstein}
\end{equation}
On Einstein–AdS manifolds, \(Y\) vanishes identically. Thus, \(Y\) measures the departure from the Einstein–AdS condition, while the AdS curvature coincides with the Weyl tensor as
\begin{equation}
F_{\mu \nu }^{\alpha \beta } |_E=W_{\mu \nu }^{\alpha \beta } \,.
\label{FWeyltensor}
\end{equation}
In the present construction, however, this condition is required to be preserved only asymptotically. Substituting Eq.~\eqref{Weyldecomposition} into Eq.~\eqref{infinitesimalWeyltensorvar} yields the transformation of the AdS curvature tensor as
\begin{equation}
\delta _{\sigma }F_{\mu \nu }^{\alpha \beta } =-2\sigma F_{\mu \nu }^{\alpha \beta }+\delta _{\sigma }Y_{\mu \nu }^{\alpha \beta } +2\sigma Y_{\mu \nu }^{\alpha \beta } \,.
\label{weylvarF1}
\end{equation}
Substituting the Weyl variation of the Schouten tensor
\begin{equation}
 \delta _{\sigma } S_{\mu \nu} = -\nabla_{\mu} \nabla_{\nu} \sigma \,,
\label{weylvarRicciSchouten}
\end{equation}
into Eq.~\eqref{Ytensor}, gives
\begin{equation}\delta _{\sigma }Y_{\mu \nu }^{\alpha \beta } =4\delta _{[\mu }^{[\alpha }\delta _{\sigma }S_{\nu ]}^{\beta ]} = -8\sigma \delta _{[\mu }^{[\alpha }S_{\nu ]}^{\beta ]} -4\delta _{[\mu }^{[\alpha } \nabla _{\nu ]} \nabla ^{\beta ]}\sigma \,,
\label{weylvarXhi}
\end{equation}
that is the Weyl transformation of the $Y$ tensor. As expected, $F_{\mu \nu }^{\alpha \beta }$ is not a Weyl covariant object for an unrestricted scalar function.

For the class of asymptotically EAdS geometries, the $Y$ tensor vanishes through the non-normalizable orders, i.e., $Y_{\mu \nu }^{\alpha \beta } \simeq 0$. Requiring this property to be preserved  under infinitesimal Weyl deformations gives
\begin{equation}
\delta _{\Sigma }Y_{\mu \nu }^{\alpha \beta } \simeq 4\delta _{[\mu }^{[\alpha }\left (\frac{\Sigma }{\ell ^{2}}\delta _{\nu ]}^{\beta ]} - \nabla _{\nu ]} \nabla ^{\beta ]}\Sigma \right )  \simeq 0
\,.
\end{equation}
This relation is satisfied for Weyl parameters $\Sigma\left(X\right)$ which are solutions of the differential equation~\eqref{weyleinsteincondition}. In this case, the $F$ tensor transforms homogeneously as
\begin{equation}
\delta _{\Sigma }F_{\mu \nu }^{\alpha \beta }  
\simeq -2\Sigma F_{\mu \nu }^{\alpha \beta }\,. 
\end{equation}

\section{Recursive relations}
\label{recursive}

\subsection{Recursion relation for $K_{\left(i\right)}^{q}$}

Using $O_{\left(i\right)}^{q+1}=\Box O_{\left(i\right)}^{q}$, the quantity $K_{\left(i\right)}^{q+1}$ is given by
\begin{equation}
K_{\left(i\right)}^{q}=\nabla^{\lambda}\Sigma \nabla_\lambda \Box O_{\left(i\right)}^{q} \,.
\end{equation}
To derive the recursion, evaluate $\Box K_{\left(i\right)}^{q}$:
\begin{equation}
\Box K_{\left(i\right)}^{q} = \Box \left(\nabla^{\lambda}\Sigma \nabla_\lambda O_{\left(i\right)}^{q}\right) = \nabla^\lambda O_{\left(i\right)}^{q} \Box \nabla_\lambda\Sigma + \nabla^\lambda \Sigma \Box \nabla_\lambda O_{\left(i\right)}^{q} + 2 \nabla^\mu \nabla^\lambda \Sigma \nabla_\mu \nabla_\lambda O_{\left(i\right)}^{q} \,.
\end{equation}
This expression can be further simplified when Eq.~\eqref{weyleinsteincondition} is considered. In particular,
\begin{align}
\Box \nabla_\lambda\Sigma &= g^{\mu \nu} \nabla_\mu \nabla_\nu \nabla_\lambda \Sigma \simeq g^{\mu \nu} \nabla_\mu \left(\frac{1}{\ell^2} g_{\nu \lambda} \Sigma\right) \Rightarrow \nonumber \\
\Rightarrow \Box \nabla_\lambda\Sigma & \simeq\frac{1}{\ell^2} \nabla_\lambda \Sigma \,.
\end{align}
Hence $\nabla_\lambda \Sigma$ satisfies the Laplacian eigenvalue equation with eigenvalue $\ell^{-2}$ to the stated asymptotic order.

Applying the Laplacian to $\nabla_\lambda O_{\left(i\right)}^{q}$ gives
\begin{align}
\square  \nabla _{\lambda } O_{\left(i\right)}^{q}  &=g^{\mu \nu } \nabla _{\mu } \nabla _{\nu } \nabla _{\lambda } O_{\left(i\right)}^{q} =g^{\mu \nu } \nabla _{\mu } \nabla _{\lambda } \nabla _{\nu } O_{\left(i\right)}^{q}  \,, \\
&\left [ \nabla _{\mu } , \nabla _{\lambda }\right ] \nabla _{\nu } O_{\left(i\right)}^{q}  =R_{\ \nu \lambda \mu }^{\kappa } \nabla _{\kappa } O_{\left(i\right)}^{q} \,,
\end{align}
such that the first line reads
\begin{equation}
g^{\mu \nu } \nabla _{\mu } \nabla _{\lambda } \nabla _{\nu } O_{\left(i\right)}^{q} = \nabla _{\lambda } \Box O_{\left(i\right)}^{q}  +R_{\lambda }^{\kappa } \nabla _{\kappa } O_{\left(i\right)}^{q}  \,.
\end{equation}
For asymptotically Einstein-AdS spacetimes, i.e., $R_{\lambda }^{\mu } \simeq -\frac{D -1}{\ell ^{2}} \delta_{\lambda }^{\mu }$, the last formula reduces to
\begin{equation}
\square  \nabla _{\lambda } O_{\left(i\right)}^{q}  \simeq \nabla _{\lambda } O_{\left(i\right)}^{q+1}  -\frac{D -1}{\ell ^{2}} \nabla _{\lambda } O_{\left(i\right)}^{q}  \,.
\end{equation}
Equivalently,
\begin{equation}
\square  \nabla _{\lambda } O_{\left(i\right)}^{q} \simeq \nabla_\lambda\left(\Box-\frac{D-1}{\ell^2}\right) O_{\left(i\right)}^{q} \,.
\end{equation}
Hence, we finally obtain
\begin{equation}
\Box K_{\left(i\right)}^{q} \simeq K_{\left(i\right)}^{q+1}-\frac{D-2}{\ell^2} K_{\left(i\right)}^{q} +\frac{2}{\ell^2}\Sigma O_{\left(i\right)}^{q+1} \,.
\label{recursiveKqapp}
\end{equation}

\subsection{Restricted variation of $O_{\left(i\right)}^{q}$}

We next derive the recursion for the restricted variation of $O_{\left(i\right)}^{q}$. Since $O_{\left(i\right)}^{q+1}=\Box O_{\left(i\right)}^{q}$, we get
\begin{equation}
\delta_\Sigma O_{\left(i\right)}^{q+1}= \delta_\Sigma \Box O_{\left(i\right)}^{q}= -2\Sigma O_{\left(i\right)}^{q+1}+ g^{\mu \nu} \delta_\Sigma \left(\nabla _\mu \nabla_\nu O_{\left(i\right)}^{q}\right) \,.
\end{equation}
The second term of the RHS can be further expanded as
\begin{equation}
\delta_\Sigma \left(\nabla _\mu \nabla_\nu O_{\left(i\right)}^{q}\right) = \nabla _\mu \nabla_\nu \delta_\Sigma O_{\left(i\right)}^{q} -\nabla_\lambda O_{\left(i\right)}^{q} \delta_\Sigma \Gamma^{\lambda}_{\mu \nu} \,,
\end{equation}
where
\begin{equation}
\delta_\Sigma \Gamma^{\lambda}_{\mu \nu}= \delta^{\lambda}_{\mu} \nabla_\nu \Sigma+\delta^{\lambda}_{\nu} \nabla_\mu \Sigma - g_{\mu \nu} \nabla^{\lambda} \Sigma \,.
\end{equation}
Therefore,
\begin{equation}
\delta_\Sigma O_{\left(i\right)}^{q+1} = -2\Sigma O_{\left(i\right)}^{q+1}+\Box \delta_\Sigma O_{\left(i\right)}^{q} +\left(D-2\right) \nabla^\lambda \Sigma \nabla_\lambda O_{\left(i\right)}^{q} \,,
\end{equation}
or equivalently
\begin{equation}
\delta_\Sigma O_{\left(i\right)}^{q+1} = -2\Sigma O_{\left(i\right)}^{q+1}+\Box \delta_\Sigma O_{\left(i\right)}^{q} +\left(D-2\right) K_{\left(i\right)}^{q} \,.
\end{equation}
A very useful generic relation in our method is the one involving $\Box \left(\Sigma O_{\left(i\right)}^{q}\right)$. In that case, applying the Laplacian on the argument gives
\begin{equation}
\Box \left(O_{\left(i\right)}^{q}\right)= O_{\left(i\right)}^{q} \Box \Sigma + \Sigma O_{\left(i\right)}^{q+1} + 2 K_{\left(i\right)}^{q} \,.
\end{equation}
Taking the trace of Eq.~\eqref{weyleinsteincondition}, the latter can be cast in the following form
\begin{equation}
\Box \left(\Sigma O_{\left(i\right)}^{q}\right) \simeq \frac{D}{\ell^2} \Sigma O_{\left(i\right)}^{q} + \Sigma O_{\left(i\right)}^{q+1}+ 2 K_{\left(i\right)}^{q} \,.
\end{equation}

\subsection{First levels of the hierarchy}

We now evaluate the first three levels explicitly in order to exhibit the triangular structure used in Sec.~\ref{closure}. For $q=0$, we get
\begin{equation}
\delta_\Sigma O_{\left(i\right)}^{1} = -2\Sigma O_{\left(i\right)}^{1}+\Box \delta_\Sigma O_{\left(i\right)}^{0} +\left(D-2\right) K_{\left(i\right)}^{0} \,,
\end{equation}
where $O_{\left(i\right)}^{0}=\Phi _{\left (i\right )}$.
Using the seed relation~\eqref{K0} together with Eq.~\eqref{boxsigmaOq} at $q=0$, we obtain that
\begin{equation}
\delta_\Sigma O_{\left(i\right)}^{1} \simeq \left(\Delta_{\left (i\right )}-2\right) \Sigma O_{\left(i\right)}^{1}+2 \left(D+\Delta_{\left (i\right )}-1\right) K_{\left(i\right)}^{0}\,.
\label{weylO1}
\end{equation}
For $q=1$, Eq.~\eqref{deltasigmaOq} reads
\begin{equation}
\delta_\Sigma O_{\left(i\right)}^{2} = -2\Sigma O_{\left(i\right)}^{2}+\Box \delta_\Sigma O_{\left(i\right)}^{1} +\left(D-2\right) K_{\left(i\right)}^{1} \,.
\end{equation}
Using the Weyl variation of $O_{\left(i\right)}^{1}$ in Eq.~\eqref{weylO1}, we obtain the form
\begin{equation}
\delta_\Sigma O_{\left(i\right)}^{2} \simeq \left(\Delta_{\left (i\right )}-4\right)\Sigma O_{\left(i\right)}^{2}+\frac{D\left(\Delta_{\left (i\right )}-2\right)}{\ell^2} \Sigma O_{\left(i\right)}^{1} +\left(D+2\Delta_{\left (i\right )}-6\right) K_{\left(i\right)}^{1} +2 \left(D+\Delta_{\left (i\right )}-1\right) \Box K_{\left(i\right)}^{0} \,.
\end{equation}
Introducing the recursive relation for $K_{\left(i\right)}^{q}$ in Eq.~\eqref{Krecursive}, we express it in the equivalent form
\begin{align}
\delta_\Sigma O_{\left(i\right)}^{2} &\simeq \left(\Delta_{\left (i\right )}-4\right)\Sigma O_{\left(i\right)}^{2}+\frac{\left(D-4\right)\Delta_{\left (i\right )} -4\left(D-3\right)}{\ell^2} \Sigma O_{\left(i\right)}^{1} +\left(3D+4\Delta_{\left (i\right )}-8\right) \Box K_{\left(i\right)}^{0}  \nonumber \\
&+\frac{\left(D-2\right) \left(D+2\Delta_{\left (i\right )}-6\right)}{\ell^2}K_{\left(i\right)}^{0}\,.
\end{align}
Finally, considering the definition of $K_{\left(i\right)}^{0}$ along the restricted Weyl orbit~\eqref{K0} and Eq.~\eqref{boxsigmaOq}, we get that
\begin{equation}
\delta_\Sigma O_{\left(i\right)}^{2} 
\simeq \left (\Delta_{\left (i\right )}  -4\right ) \Sigma O_{\left(i\right)}^{2} +\frac{4\left [\left (D -3\right )\left (\Delta_{\left (i\right )}  -1\right ) +\Delta_{\left (i\right )}^{2}\right ]}{\ell ^{2}}\Sigma O_{\left(i\right)}^{1}+\frac{4\Delta_{\left (i\right )} \left (D +\Delta_{\left (i\right )}  -1\right )\left (D +2\Delta_{\left (i\right )}  -3\right )}{\ell ^{4}}\Sigma O_{\left(i\right)}^{0} \,.
\label{weylsigmaO2}
\end{equation}
For $q=2$, the same procedure gives
\begin{gather}
\delta _{\Sigma } O_{\left(i\right)}^{3} \simeq \left (\Delta_{\left (i\right )}  -6\right )\Sigma  O_{\left(i\right)}^{3} +\frac{2\left [3D\left (\Delta_{\left (i\right )}  -2\right ) +3 \Delta_{\left (i\right )} \left (\Delta_{\left (i\right )}  -5\right ) +26\right ]}{\ell ^{2}}\Sigma O_{\left(i\right)}^{2} \nonumber  \\
+\frac{4\left [D^{2}\left (3 \Delta_{\left (i\right )}  -2\right ) +D\left (\Delta_{\left (i\right )}  -2\right )\left (9 \Delta_{\left (i\right )}  -10\right ) + \Delta_{\left (i\right )} \left (\Delta_{\left (i\right )} \left (6 \Delta_{\left (i\right )}  -31\right ) +57\right ) -42\right ]}{\ell ^{4}}\Sigma  O_{\left(i\right)}^{1} \nonumber  \\
+\frac{8 \Delta_{\left (i\right )} \left (D +\Delta_{\left (i\right )}  -1\right )\left [D\left (D -10\right ) +6 \Delta_{\left (i\right )}^{2} +6D \Delta_{\left (i\right )}  -22 \Delta_{\left (i\right )}  +21\right ]}{\ell ^{6}}\Sigma O_{\left(i\right)}^{0} \,. \label{weylvarboxcubephii}
 \end{gather} 
Multiplying these expressions by $\sqrt {\vert g \vert} $ and including the variation of the volume density reproduces the triangular transformation law~\eqref{weylBiq} through $q=3$.

\section{Curvature invariants for the ten-dimensional black holes}
\label{10Dcurvature}

For completeness, we collect here the curvature polynomials entering the ten-dimensional analysis of Sec.~\ref{topologicalBH}. The three Einstein bases are normalized according to $\widehat R_{ij}=\widehat g_{ij}$, corresponding to $\lambda=1$ in Eq.~\eqref{intrinsicbaseRicci}. For each geometry we list the invariants $\Phi_{\left(k\right)}$, with $2\leq k\leq5$, obtained by direct contraction of the AdS curvature tensor. Their Laplacian descendants are not displayed explicitly, since they follow directly by repeated application of the radial operator in Eq.~\eqref{radial-Laplacian-D2n}.

These expressions separate naturally into universal terms carrying the highest powers of the mass parameter and base-dependent contributions controlled by the intrinsic curvature of the base. The former reproduce Eq.~\eqref{massdependent10D}, while the latter distinguish the three geometries at the level of the local action density.

For \(\mathcal B_{8}=(\mathbb{S}^{2})^{4}\), the computation gives
\begin{align}
\Phi_{\left(2\right)} \left[{\left(\mathbb{S}^{2}\right)^{4}}\right]
&=\frac{384\left(r^{14}+1176 m^2\right)}{7r^{18}}
\,,
\label{eq:Phi2-D10-S2}
\\
\Phi_{\left(3\right)} \left[{\left(\mathbb{S}^{2}\right)^{4}}\right]
&= \frac{6144\left(4 r^{21}+63 m r^{14}+ 262395 m^3 \right)}{49 r^{27}} \,,
\label{eq:Phi3-D10-S2}
\\
\Phi_{\left(4\right)} \left[{\left(\mathbb{S}^{2}\right)^{4}}\right]
&=\frac{73728 \left(19 r^{28}+ 1120 m r^{21} +45864  m^2 r^{14} + 16855020 m^4\right)}{343 r^{36}}
\,,
\label{eq:Phi4-D10-S2}
\\
\Phi_{\left(5\right)} \left[{\left(\mathbb{S}^{2}\right)^{4}}\right]
&=\frac{11796480 m\left(19 r^{28} + 140 m r^{21} +1617 m^2 r^{14} +216090 m^4\right)}{49 r^{45}} \,.
\label{eq:Phi5-D10-S2}
\end{align}

For \(\mathcal B_{8}=\left(\mathbb{S}^{2}\right)^{2}\times\mathbb{CP}^{2}\), the corresponding expressions are
\begin{align}
\Phi_{\left(2\right)} \left[{\left(\mathbb{S}^{2}\right)^{2}\times\mathbb{CP}^{2}}\right]
&=
\frac{32\left(29 r^{14} + 42336 m^2\right)}{21 r^{18}} \,,
\label{eq:Phi2-D10-S2S2CP2}
\\
\Phi_{\left(3\right)} \left[{\left(\mathbb{S}^{2}\right)^{2}\times\mathbb{CP}^{2}}\right]
&=
\frac{512 \left(34 r^{21}+609 m r^{14}+ 3148740 m^3 \right)}{49 r^{27}} \,,
\label{eq:Phi3-D10-S2S2CP2}
\\
\Phi_{\left(4\right)} \left[{\left(\mathbb{S}^{2}\right)^{2}\times\mathbb{CP}^{2}}\right]
&=
\frac{4096 \left(209 r^{28}+ 14280 m r^{21} +665028  m^2 r^{14} + 303390360 m^4\right)}{343 r^{36}} \,,
\label{eq:Phi4-D10-S2S2CP2}
\\
\Phi_{\left(5\right)} \left[{\left(\mathbb{S}^{2}\right)^{2}\times\mathbb{CP}^{2}}\right]
&=
\frac{327680 m\left(418 r^{28} + 3570 m r^{21} +46893 m^2 r^{14} +7779240 m^4\right)}{49 r^{45}} \,.
\label{eq:Phi5-D10-S2S2CP2}
\end{align}

Finally, for
\(\mathcal B_{8}=\mathbb{CP}^{2}\times\mathbb{CP}^{2}\), one finds
\begin{align}
\Phi_{\left(2\right)} \left[{\mathbb{CP}^{2}\times\mathbb{CP}^{2}}\right]
&=
\frac{64\left(11 r^{14} + 21168 m^2\right)}{21 r^{18}} \,,
\label{eq:Phi2-D10-CP2CP2}
\\
\Phi_{\left(3\right)} \left[{\mathbb{CP}^{2}\times\mathbb{CP}^{2}}\right]
&=
\frac{1024 \left(10 r^{21}+231 m r^{14}+ 1574370 m^3 \right)}{49 r^{27}} \,,
\label{eq:Phi3-D10-CP2CP2}
\\
\Phi_{\left(4\right)} \left[{\mathbb{CP}^{2}\times\mathbb{CP}^{2}}\right]
&=
\frac{2048 \left(799 r^{28}+ 50400 m r^{21} +3027024  m^2 r^{14} + 1820342160 m^4\right)}{1029 r^{36}} \,,
\label{eq:Phi4-D10-CP2CP2}
\\
\Phi_{\left(5\right)} \left[{\mathbb{CP}^{2}\times\mathbb{CP}^{2}}\right]
&=
\frac{327680 m\left(799 r^{28} + 6300 m r^{21} +106722 m^2 r^{14} +23337720 m^4\right)}{147 r^{45}} \,.
\label{eq:Phi5-D10-CP2CP2}
\end{align}
The coefficients of the highest powers $m^2$, $m^3$, $m^4$, and $m^5$ are identical for all three base geometries, yielding the universal contributions displayed in Eq.~\eqref{massdependent10D}. The remaining terms depend on the intrinsic curvature of $\mathbb{B}_8$ and encode the local distinction among $(\mathbb{S}^2)^4$, $(\mathbb{S}^2)^2\times\mathbb{CP}^2$, and $\mathbb{CP}^2\times\mathbb{CP}^2$. After the complete polynomial and derivative contributions are combined, this local dependence reduces in the renormalized action to the overall volume $V_{\mathbb{B}_8}$, as shown in Sec.~\ref{topologicalBH}.

\bibliographystyle{JHEP}
\bibliography{biblioWeyl.bib}

\end{document}